\documentclass[onecolumn,11pt]{article}
\usepackage[top=1in, bottom=1in, left=1in, right=1in]{geometry}
\usepackage{amsmath,amsfonts,amscd,amssymb}
\usepackage{graphicx}
\usepackage{epstopdf}
\usepackage{overpic}
\usepackage{cancel}
\usepackage{rotating}
\usepackage{url}
\usepackage{caption}
\usepackage{color}
\usepackage{rotating}
\usepackage{multirow}
\usepackage{wrapfig}
\usepackage{mathtools}
\usepackage{subeqnarray}
\usepackage{setspace}
\usepackage{palatino} 
\usepackage[bottom,flushmargin,hang,multiple]{footmisc}
\usepackage{lipsum}
\newcommand\blfootnote[1]{%
  \begingroup
  \renewcommand\thefootnote{}\footnote{#1}%
  \addtocounter{footnote}{-1}%
  \endgroup
}

\DeclareMathOperator*{\argmin}{arg\rm{}min}

\newcommand{\bx}{\mathbf{x}}

\newcommand{\bA}{\mathbf{A}}

\newcommand{\bX}{\mathbf{X}}

\newcommand{\bphi}{\boldsymbol{\phi}}
\newcommand{\bPhi}{\boldsymbol{\Phi}}

\newcommand{\bSigma}{\boldsymbol{\Sigma}}

\usepackage{natbib}

\definecolor{offlinered}{RGB}{115, 0, 0}
\definecolor{onlineblue}{RGB}{4, 79, 149}

\title{\LARGE{\vspace{-.25in}\textbf{Robust Principal Component Analysis for\\ Modal Decomposition of Corrupt Fluid Flows}}}
\author{Isabel Scherl$^{1*}$, Benjamin Strom$^1$, Jessica K. Shang$^2$, \\ Owen Williams$^3$, Brian L. Polagye$^1$, and Steven L. Brunton$^1$\\
\footnotesize{$^1$ Department of Mechanical Engineering, University of Washington, Seattle, WA 98195, USA}\\
\footnotesize{$^2$ Department of Mechanical Engineering, University of Rochester, Rochester, NY 14627, USA}\\
\footnotesize{$^3$ Department of Aeronautics and Astronautics, University of Washington, Seattle, WA 98195, USA}
}
\date{}
\begin{document}

\maketitle

\blfootnote{$^*$ Corresponding author (ischerl@uw.edu).\\ \noindent \textbf{Matlab code:}  github.com/ischerl/RPCA-PIV \\ \noindent \textbf{Videos:}  tinyurl.com/RPCA-PIV}

\vspace{-.2in}
\begin{abstract}
Modal analysis techniques are used to identify patterns and develop reduced-order models in a variety of fluid applications. 
However, experimentally acquired flow fields may be corrupted with incorrect and missing entries, which may degrade modal decomposition. 
Here we use robust principal component analysis (RPCA) to improve the quality of flow field data by leveraging global coherent structures to identify and replace spurious data points. 
RPCA is a robust variant of principal component analysis (PCA), also known as proper orthogonal decomposition (POD) in fluids, that decomposes a data matrix into the sum of a low-rank matrix containing coherent structures and a sparse matrix of outliers and corrupt entries.  
We apply RPCA filtering to a range of fluid simulations and experiments of varying complexities and assess the accuracy of low-rank structure recovery. 
First, we analyze direct numerical simulations of flow past a circular cylinder at Reynolds number 100 with artificial outliers, alongside similar PIV measurements at Reynolds number 413.
Next, we apply RPCA filtering to a turbulent channel flow simulation from the Johns Hopkins Turbulence database, demonstrating that dominant coherent structures are preserved in the low-rank matrix. 
Finally, we investigate PIV measurements behind a two-bladed cross-flow turbine that exhibits both broadband and coherent phenomena.
In all cases, we find that RPCA filtering extracts dominant coherent structures and identifies and fills in incorrect or missing measurements.  
The performance is particularly striking when flow fields are analyzed using dynamic mode decomposition, which is sensitive to noise and outliers.  
\end{abstract}

\section{Introduction}
The ability to understand, model, and control fluid flows is foundational to advancing technologies in transportation, energy, health, and defense. 
The challenges these fields pose are not easily solved by first principles analysis without intense simplification. 
Instead, we rely on data from simulations and experiments~\citep{taira2017aiaa,Duraisamy2018arfm,Brunton2020arfm}.
The fidelity of both approaches have improved dramatically, generating vast and increasing volumes of data~\cite{Pollard2016book}. 
However, despite this apparently high-dimensional data, fluid dynamics are often characterized by the evolution of a few dominant coherent structures that are energetically or dynamically important~\cite{Lumley:1970,aubry1988dynamics,berkooz1993proper,HLBR_turb,Kutz2016book}. 
Thus, even with increasing ambient measurement dimension, the \emph{intrinsic} dimension of the flow may remain relatively low.  
Modal decomposition techniques are designed to extract these meaningful patterns from high-dimensional fluids data~\cite{taira2017aiaa,taira2019aiaa}, resulting in a compact representation that can be used for accurate and efficient reduced-order models and control~\cite{Brunton2015amr,Rowley2017arfm}.  

The majority of modal decompositions are linear~\cite{taira2017aiaa,taira2019aiaa}, although emerging techniques in machine learning are providing improved nonlinear pattern extraction~\cite{Brunton2020arfm}. 
Linear regression and least-squares optimization are particularly widely used, as in the ubiquitous proper orthogonal decomposition (POD) (also known as principal component analysis (PCA)),~\cite{Lumley:1970,aubry1988dynamics,berkooz1993proper,HLBR_turb,Towne2018jfm} and the emerging dynamic mode decomposition (DMD)~\cite{Schmid2010jfm,Rowley2009jfm,Tu2014jcd,Kutz2016book}. 
POD provides a principled approach to decomposing high-dimensional fluid flow data into a hierarchy of orthogonal modes that are ordered in terms of their ability to capture the energy in the flow; because these modes are orthogonal, it is possible to obtain reduced-order models by Galerkin projection of the Navier-Stokes equations onto a truncated POD basis~\cite{Noack2003jfm,Carlberg2017jcp,Loiseau2017jfm}. 
DMD is a related technique to decompose a flow into spatiotemporal coherent structures that are each constrained to have coherent and linear dynamics in time. 
The least-squares regression underlying both of these approaches is highly susceptible to outliers and corrupted data~\cite{candes2011jacm,Brunton2019book}, which is common in experimental measurement techniques such as particle image velocimetry (PIV).  
Thus, modal decomposition techniques such as POD/PCA and DMD are \emph{fragile} with respect to outliers.  
Further, even though POD is robust to Gaussian white noise~\cite{Gavish2014arxiv}, DMD is sensitive to noisy data~\citep{Bagheri2014pof,Dawson2016ef,Hemati2017tcfd}.   
Even techniques that are robust to corrupt velocity fields, such as finite-time Lyapunov exponents (FTLE) and Lagrangian coherent structures (LCS) \citep{Haller2002pof,shadden2005pnp,shadden2007jfm,Green2011jfm,raben2014eif,Farazmand2012chaos}, are unable to process velocity fields with missing regions, which is common in experimentally acquired data, so that interpolation must be used to fill in missing vectors. 
In this work, we explore the use of the robust principal component analysis (RPCA)~\citep{candes2011jacm} to process corrupt flow fields, leveraging global correlations in the data. 
We emphasize the impact of this approach on modal analysis, including POD/PCA and DMD. 

\subsection{Experimental challenges}
Experimental techniques to measure fluid flows have evolved rapidly over the past century, with the ultimate goal of acquiring full flow fields with high spatial and temporal resolution.
Laser-based imaging techniques have evolved from point measurements~\citep{yeh1964apl,foreman1965apl} to 2D and 3D field measurements~\citep{willert1991digital,adrian1991ar,adrian2005eif,adrian2011cup,westerweel2013ar,raffel2018particle}.  
Particle Image Velocimetry (PIV) has since become a cornerstone of experimental fluid mechanics, providing non-intrusive velocity field measurements across a range of applications.  
Improvements in PIV hardware, including more powerful lasers, higher resolution and frame rate cameras, advanced image processing technology, and the development of tomographic PIV are providing unprecedented views into real flows. 
Despite the undeniable success of PIV, there are several well-known challenges to acquiring clean and accurate data.  
Multiple factors in the PIV data acquisition and processing pipeline can contribute to velocity vector outliers that degrade the resulting velocity fields.
These include inadequate illumination and irregularities in the light field, background speckle, seeding density and non-passivity of the particles, sharp gradients in flow properties, optical issues, such as alignment and aberration, limited resolution and shot noise in the image recording, and out of plane motion of the particles when measuring in 2D~\citep{huang1997mst,adrian2011cup}. 
Because of a fundamental tradeoff between the quantity and quality of PIV data, in both space and time, researchers continue to push the resolution limits of current systems. 
Thus, flow fields acquired with PIV often have 
missing and/or corrupt measurements.  
This has motivated processing techniques to improve PIV data quality of PIV data~\citep{hart2000eif,nogueira1997data,garcia2011eif}, including predictor-corrector schemes~\citep{schrijer2008effect} and spatial filtering to remove frequencies not possible for the measurement resolution~\citep{discetti2013spatial}. 
The identification of spurious vectors has been studied extensively~\citep{westerweel1994eif,duncan2010universal}, and the normalized median filter is a robust and well-used method~\citep{westerweel2005eif}. 
However, missing vectors often cluster in regions of high shear, presenting a challenge for standard vector validation and interpolation methods that rely on local flow information~\citep{westerweel2005eif}. 
In this work, we leverage robust statistics and global spatiotemporal coherent structures across the entire dataset to fill in missing measurements and improve modal decomposition of fluid flow fields.

\begin{figure}
\vspace{.15in}
\begin{center}
\begin{overpic}[width=\textwidth]{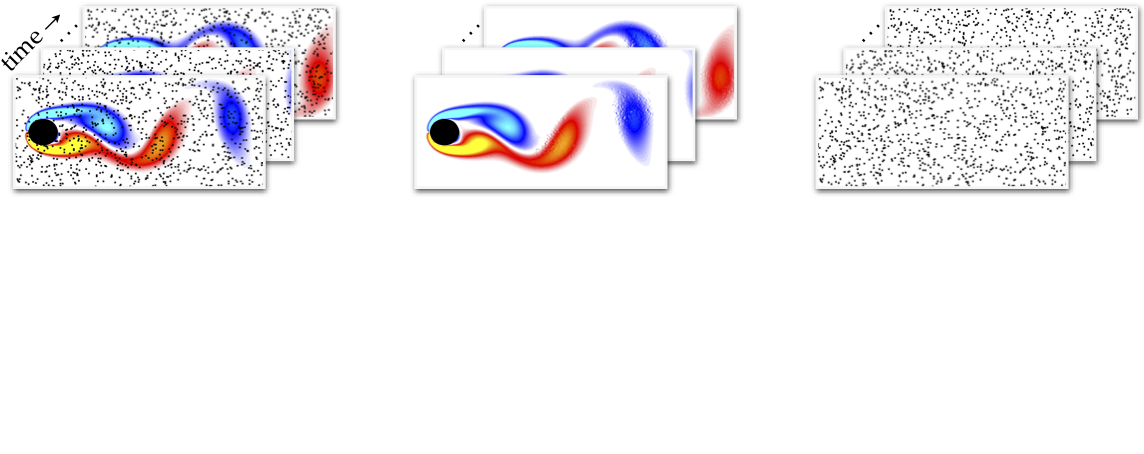}
\put(2,14){	$
	\underbrace{\begin{bmatrix}
\vline~ & \vline~ &  & \vline~~ \\
\mathbf{x}_1 & \mathbf{x}_2 & \cdots & \mathbf{x}_m \\ 
\vline~ & \vline~ & & \vline~~ 
\end{bmatrix}}_{\mathbf{X}}
\hspace{.4in}
=
\hspace{.4in}
\underbrace{\begin{bmatrix}
\vline~ & \vline~ & & \vline~~ \\
\mathbf{\tilde{x}}_1 & \mathbf{\tilde{x}}_2& \cdots & \mathbf{\tilde{x}}_m \\ 
\vline~ & \vline~ & & \vline~~
\end{bmatrix}}_{\mathbf{L}}
\hspace{.4in}
+ 
\hspace{.4in}
\underbrace{\begin{bmatrix}
\vline~ & \vline~ && \vline~~ \\
\mathbf{s}_1 & \mathbf{s}_2 &\cdots & \mathbf{s}_m \\ 
\vline~ & \vline~ && \vline~~ 
\end{bmatrix}}_{\mathbf{S}}
$	
}
\put(31.5,28.5){$=$}
\put(66,28.5){$+$}

\put(28,27.5){$\textbf{x}_m$}
\put(24.4,24){$\textbf{x}_2$}
\put(21.9,21.5){$\textbf{x}_1$}

\put(63,27){$\textbf{\~{x}}_m$}
\put(59.4,23.5){$\textbf{\~{x}}_2$}
\put(56.9,21){$\textbf{\~{x}}_1$}

\put(98.1,27.5){$\textbf{s}_m$}
\put(94.4,24){$\textbf{s}_2$}
\put(91.9,21.5){$\textbf{s}_1$}
\end{overpic}
\vspace{-.85in}
\end{center}
\caption{Schematic of RPCA filtering applied to corrupt flow field data.  Corrupted snapshots are arranged as column vectors in the matrix ${\bf X}$, which is decomposed into the sum of a low-rank matrix ${\bf L}$ and a sparse matrix of outliers ${\bf S}$. ({Videos:} {tinyurl.com/RPCA-PIV})}\label{Fig:Overview}
\end{figure}
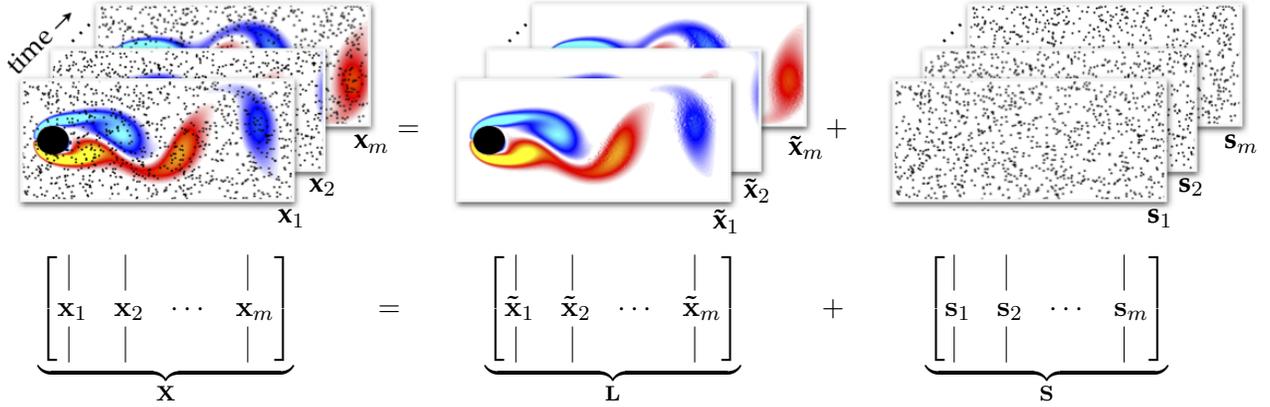

\subsection{Contributions of this work}
We investigate the use of RPCA~\citep{candes2011jacm}, a robust variant of POD/PCA~\citep{Pearson:1901,Brunton2019book}.  
RPCA uses a sparsity-promoting optimization to decompose a data matrix into the sum of a low-rank matrix containing coherent structures and a sparse matrix containing outliers.  
RPCA was originally popularized in the Netflix matrix completion algorithm for its recommender system~\citep{wong2015rad2} and has since been widely used for image and video processing~\citep{bouwmans2014cviu}, electrical capacitance tomography~\citep{lei2013sensors}, and voice separation~\citep{huang2012ieee}.  
Here, we use RPCA filtering to process flow field data from several simulations and experiments.  
 Figure \ref{Fig:Overview} demonstrates the ability of RPCA to uncover and isolate the dominant low-rank coherent structures from sparse outliers in flow data from an idealized example.   
In addition to directly analyzing and processing flow field data, we also perform PCA and DMD modal analyses on the data before and after RPCA filtering to assess its performance. 

Here we consider a range of simulated and experimentally acquired flow fields of varying complexity to isolate and analyze various aspects of the algorithm applied to data from fluid mechanics.  
First, we investigate high-fidelity flow fields from direct numerical simulations of a laminar flow past a cylinder and a turbulent channel flow, where it is possible to artificially add corrupt velocity field vectors to compare the RPCA filtered fields with a known ground truth.  
Next, we apply the method to two experimentally acquired datasets, including a companion laminar flow past a cylinder and measurements of a cross-flow turbine wake.  
Although there is not a ground truth model for these flows, it is possible to estimate the effect of RPCA filtering in reducing outliers and corruption by analyzing the DMD spectrum, which has well-stereotyped behavior for such periodic wake flows~\cite{Bagheri2014pof}.  In all cases, we show that the RPCA filtered fields yield DMD spectra that are more consistent with a periodic wake in the absence of noise.  

This work is organized as follows:  First, we present the standard POD/PCA and DMD modal analysis techniques in Sec.~\ref{Sec:Modal}, followed by the RPCA method in Sec.~\ref{Sec:RPCA}.  Section~\ref{Sec:Models} describes the four flow fields used in this analysis.  Results of RPCA filtering on these flow fields and its impact on downstream modal analysis are presented in Sec.~\ref{Sec:Results}.  

\section{Modal analysis}\label{Sec:Modal}
Extracting coherent structures from high-dimensional data has been a central challenge in fluid mechanics for decades.  
Here we review two leading modal decomposition techniques for data from fluid mechanics, the proper orthogonal decomposition (POD), also known as principal component analysis (PCA) (Sec.~\ref{Sec:POD}), and the dynamic mode decomposition (DMD) (Sec.~\ref{Sec:DMD}). 
Both methods apply equally well to data from simulations or experiments. 
We use these two modal decomposition techniques to assess the effectiveness of RPCA filtering, in processing and correcting corrupt flow fields.  
Both techniques are based on the singular value decomposition (SVD)~\cite{Golub1980siamjna,Golub1970nm,Businger1969cacm,Golub1965siamb} and there are several detailed discussions of these decompositions~\cite{Kutz2016book,taira2017aiaa,taira2019aiaa,Brunton2019book}. The RPCA algorithm is explained in Sec.~\ref{Sec:RPCA}.

\begin{figure}
\begin{center}
\begin{overpic}[width=\textwidth]{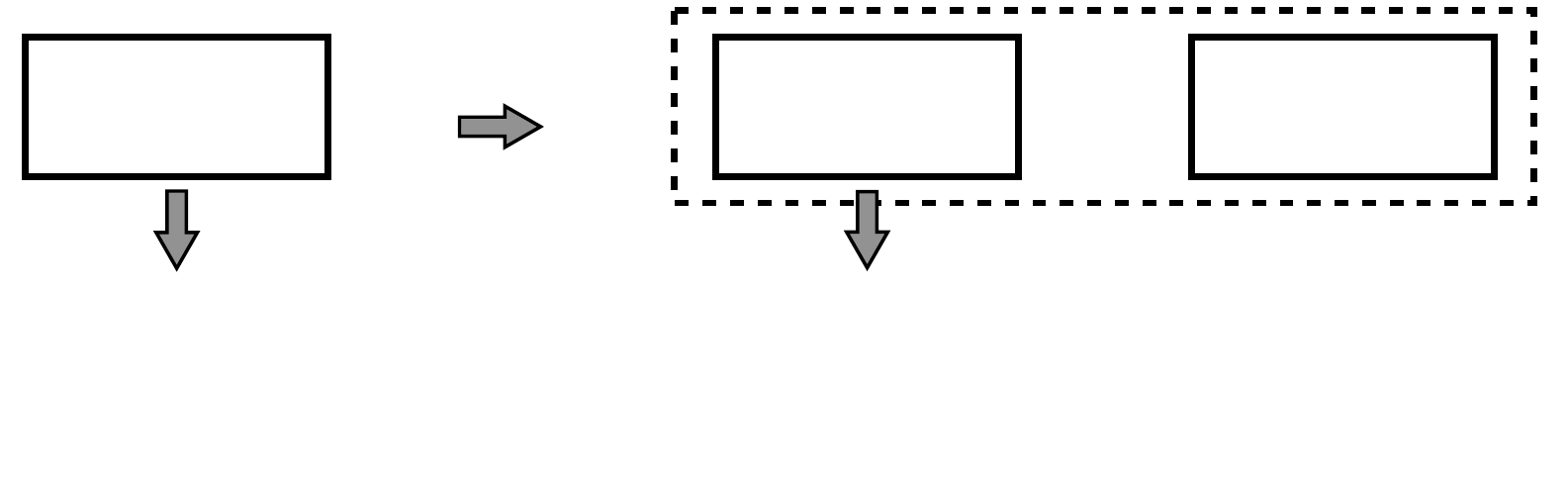}

\put(10.3,24.5){\LARGE{\textbf{X}}}
\put(5,21.5){\textbf{data matrix}}

\put(54.2,24.5){\LARGE{\textbf{L}}}
\put(51,21.5){\textbf{low rank}}

\put(85,24.5){\LARGE{\textbf{S}}}
\put(83,21.5){\textbf{sparse}}

\put(68.5,23){\LARGE{ \textbf{+}}}

\put(25,26.2){\textbf{RPCA filtering}}

\put(0.75,10.5){\textbf{PCA modes (Sec.~\ref{Sec:POD})}}
\put(0.25,8.25){\textbf{DMD modes (Sec.~\ref{Sec:DMD})}}

\put(44,10.5){\textbf{RPCA modes (Sec.~\ref{Sec:POD})}}
\put(43.5,8.25){\textbf{RDMD modes (Sec.~\ref{Sec:DMD})}}

\end{overpic}
\vspace{-.75in}
\caption{Flowchart showing how we apply RPCA filtering to a data matrix and analyze the results. Depending on the dataset in question, the data matrix may be artificially corrupted prior to RPCA filtering (Sec. \ref{Sec:Models}). The results of principal component analysis and dynamic mode decomposition performed on the data matrix (${\bf X}$) are referred to as PCA and DMD modes, respectively, wheras those same operations performed on the low rank matrix (${\bf L}$) are referred to as RPCA and RDMD modes.}\label{Fig:FlowChart}
\end{center}
\vspace{-.15in}
\end{figure}
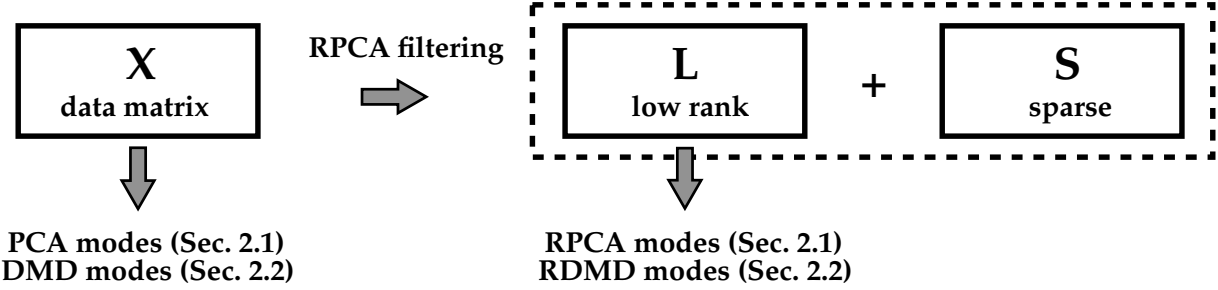

In this work, we follow the flowchart shown in Fig. ~\ref{Fig:FlowChart}. RPCA filtering is applied to a data matrix ($\textbf{X}$) which is bisected into the low rank structure ($\textbf{L}$) and sparse ($\textbf{S}$) subspaces. From there, RPCA modes and DMD modes are calculated from the low rank data. We also calculate the POD/PCA and DMD modes on the data matrix.  

\subsection{Proper orthogonal decomposition (POD)}\label{Sec:POD}
Proper orthogonal decomposition--referred to as principal component analysis (PCA) throughout results---is a widely used method to identify spatially correlated coherent structures from data, decomposing the flow field into a linear combination of orthogonal modes that are arranged hierarchically by energy content.   
There are several variants of POD~\cite{taira2017aiaa,Lumley:1970,Sirovich:1987,berkooz1993proper,HLBR_turb,Towne2018jfm}, and we will present a variant of the \emph{snapshot POD} of Sirovich that relies on the numerically stable SVD~\cite{Sirovich:1987,Brunton2019book}.  
First, flow field data (e.g., a velocity or vorticity field) is measured or computed on a discrete spatial grid, and $m$ {snapshots} of these flow fields are collected at various times $t_1,t_2, \cdots, t_m$.  
The flow field data at time $t_k$ may be reshaped into a column vector ${\bf x}_k = {\bf x}(t_k)\in\mathbb{R}^n$, where $n$ denotes the number of flow variables times the number of spatial grid locations.  
Next, a data matrix is formed by arranging the column vectors ${\bf x}_k$ in a matrix ${\bf X}$:
\begin{align}\label{Eq:DATA}
\mathbf{X}&= 
\begin{bmatrix}
\vline~ & \vline~ &  & \vline~~ \\
\mathbf{x}_1 & \mathbf{x}_2 & \cdots & \mathbf{x}_m \\ 
\vline~ & \vline~ & & \vline~~ 
\end{bmatrix}.
\end{align}
Finally, POD modes are obtained by computing the singular value decomposition of ${{\bf X}\in\mathbb{R}^{n\times m}}$:
\begin{align}\label{Eq:SVD}
{\bf X}&={\bf U}\boldsymbol{\Sigma}{\bf V}^T,
\end{align}
where $^T$ defines the matrix transpose, ${\bf U}\in {\mathbb{R}}^{n\times n}$, ${\bf \Sigma}\in {\mathbb{R}}^{n\times m}$ and ${\bf V}\in {\mathbb{R}}^{m\times m}$.   
The columns of ${\bf U}$ are \emph{POD modes} with the same dimension as a flow field ${\bf x}$. 
POD modes are orthonormal so that ${\bf U}^T{\bf U} = {\bf I}$; similarly ${\bf V}^T{\bf V} = {\bf I}$.  
Moreover, the columns of ${\bf U}$ (resp. rows of ${\bf V}^T$) are arranged in order of their importance in describing the data. 
The importance of each mode (i.e., column of ${\bf U}$) is given by the corresponding entry of the non-negative, diagonal matrix of singular values $\bSigma\in\mathbb{R}^{n\times m}$.  

The matrix ${\bf X}$ will exhibit \emph{low-rank structure}, so that it is well approximated by the first $r\ll m<n$ columns of ${\bf U}$ and ${\bf V}$:
\begin{align}\label{Eq:SVD2}
{\bf X}&\approx{\bf U}_r\boldsymbol{\Sigma}_r{\bf V}_r^T,
\end{align}
where ${\bf U}_r$ and ${\bf V}_r$ denote the first $r$ columns of each matrix and $\boldsymbol{\Sigma}_r$ denotes the first $r\times r$ sub-block of  $\boldsymbol{\Sigma}$.  
In fact, the Eckart-Young theorem states that this is the \emph{optimal} rank-$r$ approximation of the matrix ${\bf X}$ in a least-squares sense.  
More details about the SVD can be found in~\cite{Brunton2019book}. 

After truncating all but the first $r$ dominant modes, a flow field snapshot ${\bf x}$ may be approximated by a linear combination of these modes:
\begin{align*}
{\bf x} \approx \sum_{k=1}^r {\bf u}_k\alpha_k,
\end{align*}
where $\alpha_k$ is the \emph{POD mode coefficient}.  

Because the POD modal basis is orthogonal, it is possible to obtain a reduced-order nonlinear dynamical system for the evolution of the coefficients $\alpha_k(t)$ in time via Galerkin projection of the Navier-Stokes equations onto the POD basis. 
In this way, the POD basis may be thought of a data-driven generalization of the Fourier basis that is tailored to a particular flow field.  
POD is also closely related to principal component analysis (PCA)~\cite{Pearson:1901}, the Karhunen--Lo\`eve decomposition~\cite{karhunen1947lineare}, empirical orthogonal functions~\cite{lorenzMITTR56}, or the Hotelling transform~\cite{hotellingJEdPsy33_1,hotellingJEdPsy33_2}.

\subsection{Dynamic mode decomposition}\label{Sec:DMD}
DMD is a modal decomposition technique that simultaneously identifies spatially coherent modes that are constrained to have the same linear behavior in time, given by oscillations at a fixed frequency with growth or decay~\cite{Schmid2010jfm,Kutz2016book}.  
Thus, the dynamic mode decomposition provides a dimensionality reduction into a set of spatial modes along with a linear model for how these modes evolve in time.  
This is in contrast to POD, which results in orthogonal modes arranged in terms of energy content and without consideration of dynamics.  
However, in many formulations, DMD is closely related to POD, and may be thought of as a linear combination of POD modes that results in linear evolution in time.  DMD also has deep connections to nonlinear dynamical systems via Koopman operator theory~\cite{Mezic2005nd,Rowley2009jfm,Tu2014jcd,Mezic2013arfm,Kutz2016book}.

In the original formulation of DMD~\cite{Schmid2010jfm}, the snapshots in the data matrix in Eq.~\eqref{Eq:DATA} are spaced evenly in time, so that $t_k=k\Delta t$ with $\Delta t$ sufficiently small to resolve the highest frequencies in the dynamics.  
Generalizations have since been formulated to allow for non-sequential time-series~\cite{Tu2014jcd,Askham2018siads} and for data that is under-resolved in space~\cite{Brunton2015jcd,Gueniat2015pof} or time~\cite{Tu2014eif}; however, for simplicity, we will present the standard \emph{exact DMD} formulation of Tu et al.~\cite{Tu2014jcd} with evenly spaced and sequential snapshots.  
DMD seeks to identify the leading eigenvalues and eigenvectors of the best-fit linear operator ${\bf A}$ that evolves snapshots forward in time:
\begin{align}
\bx_{k+1} \approx \bA \bx_k.\label{Eq:DMD:Propagator}
\end{align}
The eigenvectors $\boldsymbol{\phi}$ of $\bA$ have the dimensions of a flow-field and correspond to spatiotemporal coherent structures whose dynamics in time evolve according to the associated eigenvalue $\gamma$.  

In practice, this operator is identified by first splitting the data in Eq.~\eqref{Eq:DATA} into two matrices:
\begin{subequations}
\begin{align}
\bX &= \begin{bmatrix} \vline & \vline & & \vline \\ 
\bx_1 & \bx_2 & \cdots & \bx_{m-1} \\
 \vline & \vline & & \vline 
 \end{bmatrix}
&\bX' &= \begin{bmatrix} \vline & \vline & & \vline \\ 
\bx_2 & \bx_3 & \cdots & \bx_m \\
 \vline & \vline & & \vline 
 \end{bmatrix}.
\end{align}
\end{subequations}
and then solving for the best fit operator that satisfies 
\begin{align}
\bX' \approx \bA \bX
\end{align}
 via the following \emph{least-squares} optimization problem
\begin{align}
\bA = \argmin_{\bA} \|\bX' - \bA \bX\|_F = \bX'\bX^\dagger \approx {\bf X}'{\bf V}_r\boldsymbol{\Sigma}_r^{-1}{\bf U}_r^T.\label{Eq:DMD:Definition}
\end{align}
Here we are minimizing the Frobenius norm  $\|\cdot\|_F$ via the pseudo-inverse ${{\bf X}^\dagger={\bf V}\boldsymbol{\Sigma}^{-1}{\bf U}^T\approx {\bf V}_r\boldsymbol{\Sigma}_r^{-1}{\bf U}_r^T}$.  

In practice, the matrix ${\bf A}$ is far too large to analyze directly, and instead, we project ${\bf A}$ onto an $r$-dimensional POD subspace, given by the columns of ${\bf U}_r$:
\begin{align}
\tilde{\bf A} = {\bf U}_r^T {\bf A}{\bf U}_r = {\bf U}_r^T{\bf X}'{\bf V}_r \boldsymbol{\Sigma}_r^{-1}.
\end{align}
The eigenvalues of $\tilde{\bf A}$ are the same as the eigenvalues of ${\bf A}$, which are known as the \emph{DMD eigenvalues}.  They are computed via the eigendecomposition of the $r\times r$ matrix $\tilde{\bf A}$:
\begin{align}
\tilde{\bf A}{\bf W} = {\bf W}\boldsymbol{\Gamma}.
\end{align}
Finally, the corresponding \emph{DMD modes} are reconstructed using the full-dimensional data along with the reduced eigenvectors in ${\bf W}$:
\begin{align}
\boldsymbol{\Phi} = {\bf X}' {\bf V}_r \boldsymbol{\Sigma}_r^{-1}{\bf W}.
\end{align}
This formula for the eigenvectors is from the exact DMD algorithm~\cite{Tu2014jcd,Kutz2016book}; the original formulation of Schmid~\cite{Schmid2010jfm} computes modes as $\boldsymbol{\Phi} = {\bf U}_r{\bf W}$.  

With the DMD modes $\bPhi$ and eigenvalues $\boldsymbol{\Gamma}$ it is possible to reconstruct the state at time $k\Delta t$
\begin{align}
\bx_k = \sum_{j=1}^r \bphi_j \gamma^{k-1} b_j = \bPhi \boldsymbol{\Gamma}^{k-1} {\bf b},
\end{align}
where the vector ${\bf b}$ of mode amplitudes is generally computed as
\begin{align}
{\bf b} = \bPhi^{\dagger}\bx_1.\label{Eq:DMDModeAmplitude}
\end{align}
More principled approaches to select the few dominant modes have been considered based on sparsity-promoting optimization~\cite{Jovanovic2014pof}.  

The spectral expansion above may also be written in continuous-time by introducing the continuous eigenvalues $\omega = \log(\gamma)/\Delta t$:
\begin{align}
\bx(t) = \sum_{j=1}^r \bphi_j e^{\omega_j t} b_j = \bPhi \exp(\boldsymbol{\Omega} t){\bf b},
\end{align}
where $\boldsymbol{\Omega}$ is a diagonal matrix containing the continuous-time eigenvalues $\omega_j$.

DMD is known to be extremely sensitive to noisy data~\citep{Bagheri2014pof,Dawson2016ef,Hemati2017tcfd}, and the eigenvalues specifically suffer from a bias that is not reduced with increasing data.  
There are several modifications to make DMD more robust to noise, including averaging forward-time and backward-time operators~\cite{Dawson2016ef}, total least squares~\cite{Hemati2017tcfd}, and variable projection~\cite{Askham2018siads}.  
For periodic wake data, as explored in three of the examples in this paper, the discrete-time eigenvalues should occur in complex conjugate pairs $\gamma,\bar\gamma$ exactly on the unit circle in the complex plane for clean data. 
Similarly, the continuous-time eigenvalues should be in complex conjugate pairs $\pm i\omega$ on the imaginary axis, indicating pure oscillations with no growth or decay~\cite{Bagheri2013jfm,Bagheri2014pof}.  
In~\cite{Bagheri2014pof}, Bagheri characterized the perturbative effect of noise on these eigenvalues, deriving an asymptotic expression for how high frequency eigenvalues become increasingly affected by noise.  If the true continuous-time eigenvalue should be $\pm i\omega$ in the absence of noise, Bagheri showed that in the presence of perturbative white noise with magnitude $\epsilon\ll 1$ the observed eigenvalue pair is 
\begin{align}\label{Eq:Shervin}
\pm i\omega - \epsilon C \omega^2 + \mathcal{O}(\epsilon^2),
\end{align}
 where $C$ is a sensitivity constant.  
 Thus, low noise levels cause a spurious real-valued damping $-\epsilon C\omega^2$ that is quadratic in the frequency. 
We will make extensive use of this property to assess the quality of our RPCA filtered fields by computing the ratio of the best-fit factor $\epsilon C$ before and after applying RPCA filtering.  
$\epsilon C$ should decrease as a consequence of reduced noise and corruption. 

\section{Robust extraction of fluid coherent structures}\label{Sec:RPCA}
Techniques based on least-squares regression, such as POD/PCA and DMD, are highly susceptible to outliers and corrupted data, making them \emph{fragile} with respect to some experimental measurement errors.  
To mitigate this sensitivity, Candes et al.~\cite{candes2011jacm} have developed a robust principal component analysis (RPCA) that seeks to decompose a data matrix ${\bf X}$ into a structured low-rank matrix ${\bf L}$ that is characterized by dominant coherent structures and a sparse matrix ${\bf S}$ containing outliers and corrupt data:
\begin{equation}
{\bf X} = {\bf L} + {\bf S}.
\end{equation}
The principal components of ${\bf L}$ are \emph{robust} to outliers and corrupt data, which are isolated in ${\bf S}$.  
This decomposition, also referred to as a filter, has profound implications for many modern problems of interest, including video surveillance (where the background objects appear in ${\bf L}$ and foreground objects appear in ${\bf S}$), facial recognition (eigenfaces are in ${\bf L}$ and shadows, occlusions, etc. are in ${\bf S}$), natural language processing and latent semantic indexing, and ranking problems\footnote{The ranking problem may be thought of in terms of the Netflix prize for matrix completion.  In the Netflix prize, a large matrix of preferences is constructed, with rows corresponding to users and columns corresponding to movies.  This matrix is sparse, as most users only rate a handful of movies.  The Netflix prize seeks to accurately fill in the missing entries of the matrix, revealing the likely user rating for movies the user has not seen.}.  

Mathematically, the goal is to find ${\bf L}$ and ${\bf S}$ that satisfy the following:
\begin{equation}
\min_{{\bf L},{\bf S}}\text{rank}({\bf L}) + \|{\bf S}\|_0 \,\,\,  \text{subject to} \,\,\, {\bf L} + {\bf S} = {\bf X}.\label{Eq:RPCA:L0}
\end{equation}
However, neither the $\text{rank}({\bf L})$ nor the $\|{\bf S}\|_0$ terms are convex, making this optimization intractable. 
Similar to the compressed sensing problem, it is possible to solve for ${\bf L}$ and ${\bf S}$ with \emph{high probability} using a convex relaxation of \eqref{Eq:RPCA:L0}:
\begin{equation}
\min_{{\bf L},{\bf S}}\|{\bf L}\|_* + \lambda_0\|{\bf S}\|_1 \,\,\,  \text{subject to} \,\,\, {\bf L} + {\bf S} = {\bf X},\label{Eq:RPCA:L1}
\end{equation}
where $\|\cdot\|_*$ is the nuclear norm, given by the sum of singular values which is a proxy for the rank of the matrix and ${\lambda_0=\lambda/\sqrt{\max(n,m)}}$ and $\|\cdot\|_1$ is the 1-norm of the matrix. 
The solution to \eqref{Eq:RPCA:L1} converges to the solution of \eqref{Eq:RPCA:L0} with high probability if $\lambda=1$, where $n$ and $m$ are the dimensions of ${\bf X}$, given that ${\bf L}$ is not sparse and ${\bf S}$ is not low-rank\index{low-rank}. 
In the examples below, these assumptions may only be partially valid, so the optimal value of $\lambda$ may vary slightly. 
The convex problem in \eqref{Eq:RPCA:L1} is known as \emph{principal component pursuit} (PCP), and may be solved using the augmented Lagrange multiplier (ALM) algorithm.  
Specifically, an augmented Lagrangian may be constructed:
\begin{equation}
\mathcal{L}({\bf L},{\bf S},{\bf Y}) = \|{\bf L}\|_* + \lambda_0\|{\bf S}\|_1 + \langle{\bf Y},{\bf X}-{\bf L}-{\bf S}\rangle + \frac{\upsilon}{2}\|{\bf X}-{\bf L}-{\bf S}\|_F^2.
\end{equation}
Where $Y$ is the matrix of Lagrange multipliers and $\upsilon$ is a hyperparameter. 
We then solve for ${\bf L}_k$ and ${\bf S}_k$ to minimize $\mathcal{L}$, update the Lagrange multipliers 
\begin{align*}
{\bf Y}_{k+1} = {\bf Y}_k + \upsilon({\bf X}-{\bf L}_k-{\bf S}_k),
\end{align*}
and iterate until convergence.  In this work, an inexact ALM implementation from~\cite{sobral2015lrs} is used.
The alternating directions method (ADM)~\citep{Lin2010adm,Yuan2009adm} provides another simple procedure.  

After the low-rank matrix ${\bf L}$ is obtained, it is possible to compute robust POD/PCA modes (Sec. \ref{Sec:POD}) as in Eq.~\eqref{Eq:SVD}:
\begin{align}
{\bf L}&={\bf U}\boldsymbol{\Sigma}{\bf V}^T.
\end{align}

Henceforth, we refer to the modes in ${\bf U}$ from ${\bf UL}$ as RPCA modes. Similarly, it is also possible to compute robust DMD (RDMD) modes and eigenvalues.

\section{Model flows}\label{Sec:Models}
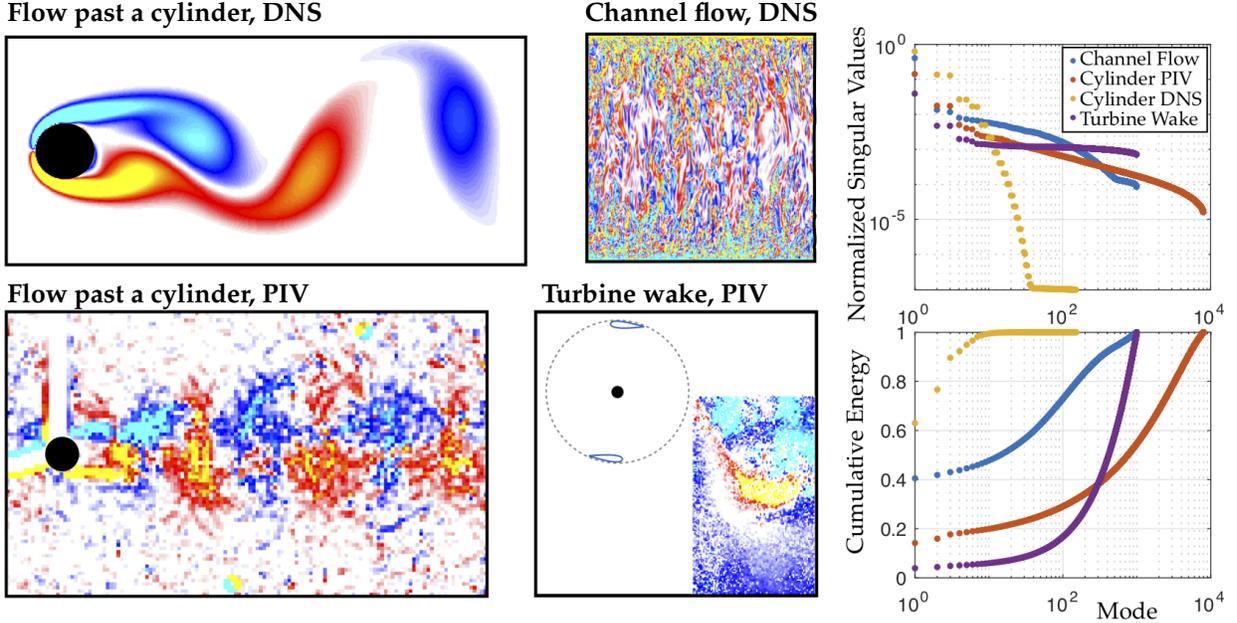
\begin{figure}
\begin{center}
\begin{overpic}[width=\textwidth]{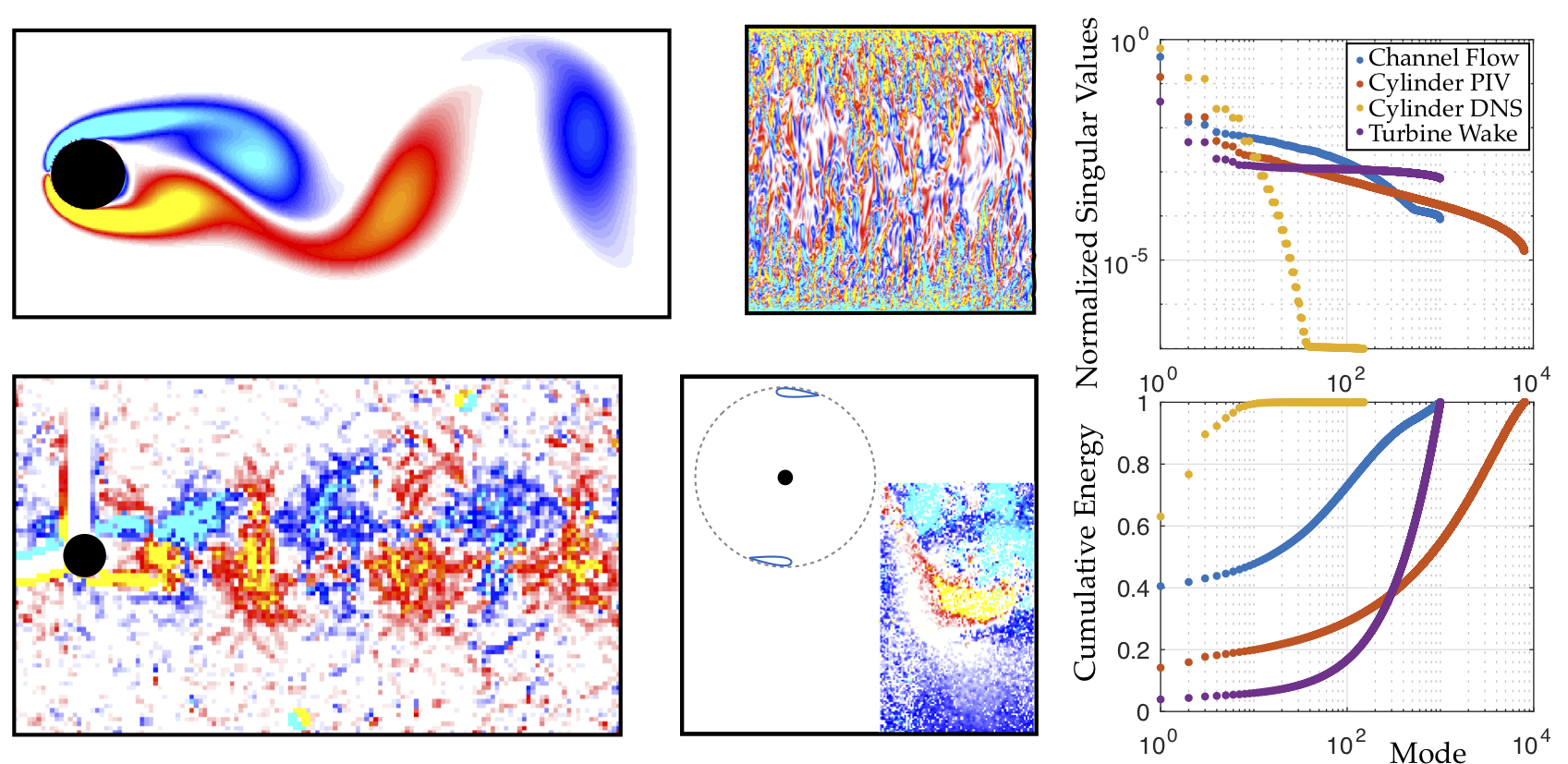}
\small
\put(1,48.2){\textbf{\small Flow past a cylinder, DNS}}
\put(1,25.5){\textbf{\small Flow past a cylinder, PIV}}
\put(44,25.5){\textbf{\small Turbine wake, PIV}}
\put(47.5,48.2){\textbf{\small Channel flow, DNS}}
\end{overpic}
\vspace{-.15in}
\caption{(left) Example flow field data.  (right) Singular value spectrum for each data set. Mean flow travels from left to right in all cases.}\label{Fig:ModelProblems}
\end{center}
\vspace{-.15in}
\end{figure}

We demonstrate RPCA filtering on several example data sets of varying complexity, drawn from direct numerical simulations (DNS) and PIV data from experiments.  
Figure~\ref{Fig:ModelProblems} provides an overview of the four example flow fields.  

\subsection{Cylinder flow} 
Flow past a cylinder is a canonical example in fluid mechanics.  
We consider data from DNS at a diameter-based Reynolds number of 100 and from PIV measurements at Reynolds number 413~\citep{Shang2015flexibility}.  

The DNS data is generated by simulating the two-dimensional incompressible Navier-Stokes equations 
using the immersed boundary projection method~\cite{taira:fastIBPM,taira:07ibfs}.  
The computational domain comprises four nested grids: the finest grid covers a domain of $9\times 4$ and the largest grid covers a domain of $72\times 32$, where lengths are non-dimensionalized by the cylinder diameter.  
Each grid contains $449\times 199$ points with a resolution of 50 points per cylinder diameter.
The time-step is $\Delta t = 0.02$ and data is sampled at intervals of $10\Delta t$ (30 times the vortex shedding frequency) with $m=150$ snapshots saved, covering $5$ vortex shedding cycles.  
The DNS provides a benchmark, where the uncorrupted flow field is known, to quantitatively assess performance of RPCA filtering on data with artificial salt \& pepper corruption.  
Corrupted sample points are chosen uniformly in space and time at a given rate, and both the $u$ and $v$ velocity components at each selected location are randomly assigned a value of $\pm10$ times the standard deviation of the streamwise velocity data.  
In addition, we consider a second case where corrupted sample points are chosen with a bias towards regions of high vorticity or shear magnitude, which is more physically realistic for PIV data.  
In this second case, we select measurements for corruption based on a probability density given by $\alpha+|\boldsymbol{\omega}|$, where $\alpha$ is a small positive constant and $|\boldsymbol{\omega}|$ is the absolute value of the vorticity; when the rate of corruption is sufficiently high, these corrupted fields begin to resemble the uniformly corrupted cases, but with more corruption in vortex cores.  
Because vorticity is calculated from velocity fields using a finite-difference derivative, there is a higher rate of corruption in the vorticity fields than in the velocity fields.

The PIV data has frame size of $135\times 80$ grid points with a resolution of $8$ points per cylinder diameter. 
Data is sampled at a rate of $20\,$Hz (125 times the shedding frequency) with $m=8,000$ snapshots saved, which corresponds to $64$ vortex shedding cycles.



\subsection{Turbulent channel flow}  
For a more complex and multi-scale flow, we consider DNS data from a forced, fully developed turbulent channel flow data with a friction velocity Reynolds number of $Re_{\tau}=1,000$, from the Johns Hopkins Turbulence Database (JHTDS)~\citep{graham2016jot}. 
This example provides a test case to see how turbulent kinetic energy at various scales is filtered depending on the level of added noise. 
The addition of noise is similar to the cylinder DNS where randomly selected sample points of the streamwise and cross-stream velocity fields are assigned a value of $\pm10$ standard deviations of the streamwise velocity data. 
Due to the size of the full dataset, we only consider two-dimensional fields on the mid-plane, with a $512\times 512$ grid of three component velocity measurements spanning the channel width. 
Data is sampled at a rate of 966 times the mean flow-through time with $m=1000$ snapshots.      

\subsection{Cross-flow turbine wake} 

Finally, we consider PIV wake data from a cross-flow turbine experiment conducted at the University of Washington. Cross-flow turbines can be used to extract power from wind and water currents for renewable energy generation.
This flow exhibits both coherent and broadband phenomena, and provides a challenging test-case RPCA filtering.  
The frame consists of $158\times 98$ grid points, at a resolution of 99 points per rotor diameter. 
Data is sampled at a rate of 32 times the blade-pass frequency with $m=1000$ snapshots. 
Vectors were calculated using a multi-grid, multi-pass algorithm with adaptive image deformation~\citep{scarano2001mst}.  
Resulting vector fields were then validated using a normalized median filter with potential replacement by secondary correlation peaks. The cross-correlation and validation steps result in missing data, particularly in regions of high vorticity and shear. 
To apply RPCA filtering, these missing values are randomly assigned a value of $\pm10$ standard deviations of the streamwise flow data, in contrast to the experimental cylinder wake where missing measurements were previously interpolated.


\section{Results}\label{Sec:Results}
We now explore the ability of RPCA filtering to isolate and remove noise and corruption from the example flow fields.  We will begin with the simulated and experimental flow past a cylinder, followed by data from the Johns Hopkins turbulent channel flow simulation, and ending with the experimental wake of a cross-flow turbine.

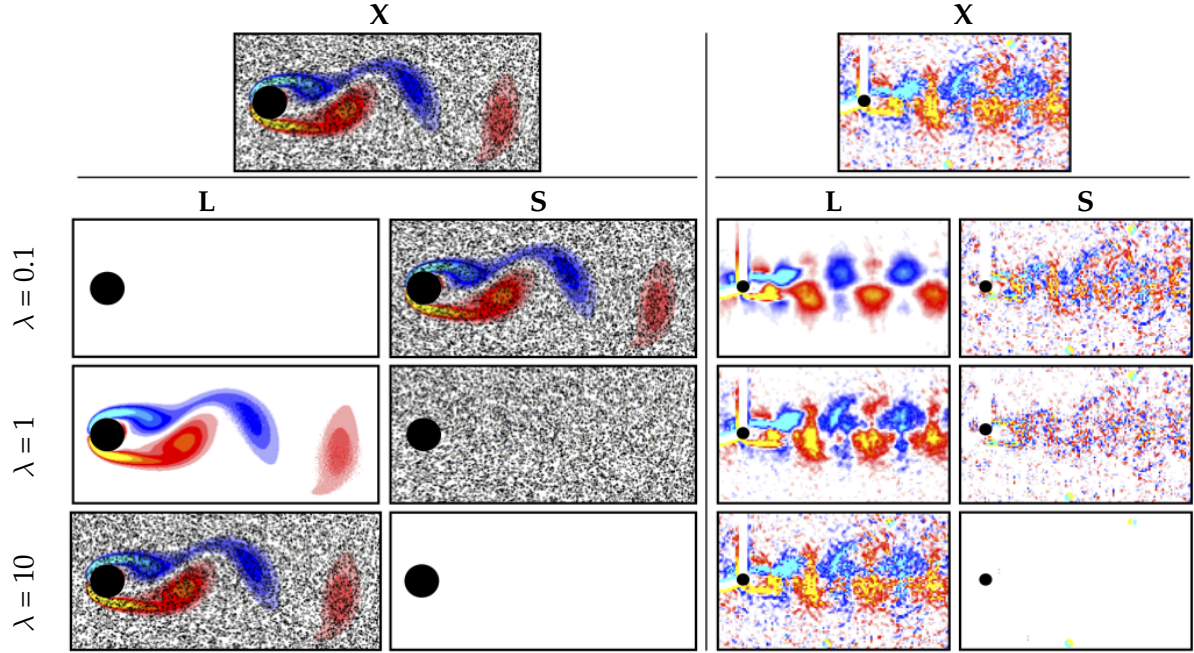
\begin{figure}
\begin{center}
\begin{overpic}[width=.99\textwidth]{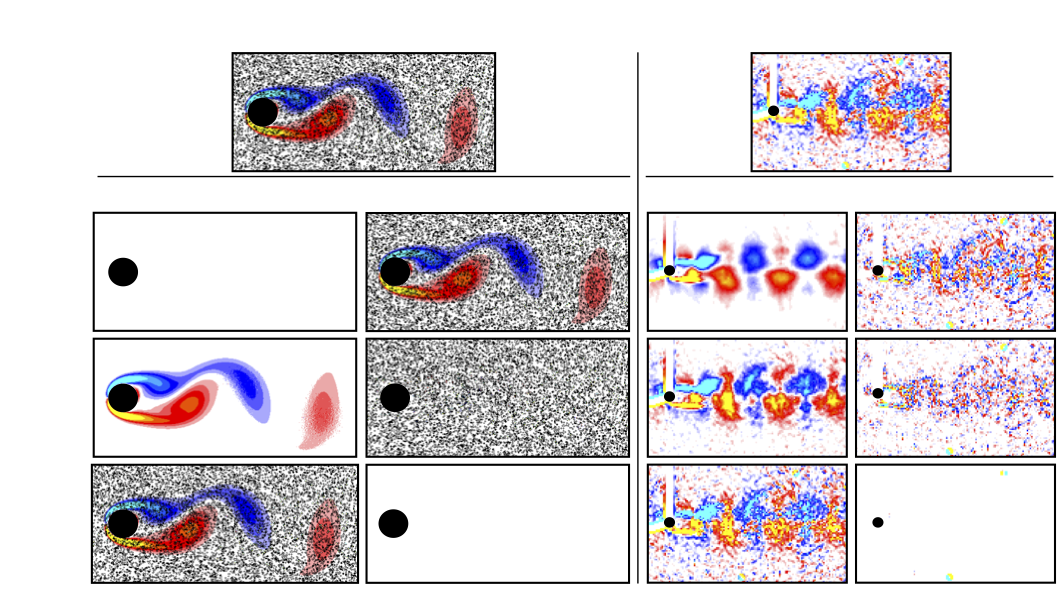}
\put(33,51.5){\textbf{X}}
\put(80.5,51.5){\textbf{X}}
\put(19,36.25){\textbf{L}}
\put(70,36.25){\textbf{L}}
\put(46,36.25){\textbf{S}}
\put(90.5,36.25){\textbf{S}}

\put(4,14.5){\begin{sideways}\textbf{$\lambda$} = 1\end{sideways}}
\put(4,26.5){\begin{sideways}\textbf{$\lambda$} = 0.1\end{sideways}}
\put(4,2){\begin{sideways}\textbf{$\lambda$} = 10\end{sideways}}
\end{overpic}
\caption{RPCA filtering removes noise and outliers in the flow past a cylinder (black circle), from DNS (left) with $10\%$ of velocity field measurements corrupted with salt and pepper noise, and PIV measurements (right). All frames show resultant vorticity fields.  As the parameter $\lambda$ is decreased, RPCA filtering is more aggressive, eventually incorrectly identifying coherent flow structures as \emph{outliers}.}\label{Fig:CylResults}
\end{center}
\end{figure}

\subsection{Cylinder flow} 
Figure~\ref{Fig:CylResults} shows the results of RPCA filtering for flow past a cylinder, providing a side-by-side comparison of PIV and corrupted DNS data data.  Although the Reynolds numbers differ by a factor of four, the flow fields are qualitatively similar, characterized by periodic, laminar vortex shedding.  
For $\lambda=1$, the data is correctly segmented with the coherent flow in ${\bf L}$ and the sparse corruption in ${\bf S}$.  
When $\lambda$ is too small, RPCA filtering is overly aggressive, incorrectly including relevant flow structures in ${\bf S}$, and when $\lambda$ is too large, the corruption is not filtered.  

For the experimental data in Fig.~\ref{Fig:CylResults} (right), the optimal value of $\lambda$ is less clear.  
For $\lambda=0.1$, the low-rank field ${\bf L}$ is visually smoother than the field at $\lambda=1$, but the sparse matrix ${\bf S}$ contains a significant portion of the wake structures, indicating over-filtering. 
This filtering becomes more pronounced in the movies, where it is clear that much of the high-frequency ``noise" in the bypass flow is actually free-stream turbulence, which is consistent with the turbulence intensity of the experiments.  
Further, as subsequently discussed, when we compute the RPCA modes, it is clear that the $\lambda=0.1$ case is heavily filtering out all but the first three modes.  
Thus, it appears that the theoretically optimal value $\lambda=1$ has the best performance, although there may be a tradeoff between filtering ambient free-stream turbulence and coherent structures of interest in experiments. 

\begin{figure}
\vspace{-.095in}
\begin{center}
\begin{overpic}[width=.934\textwidth]{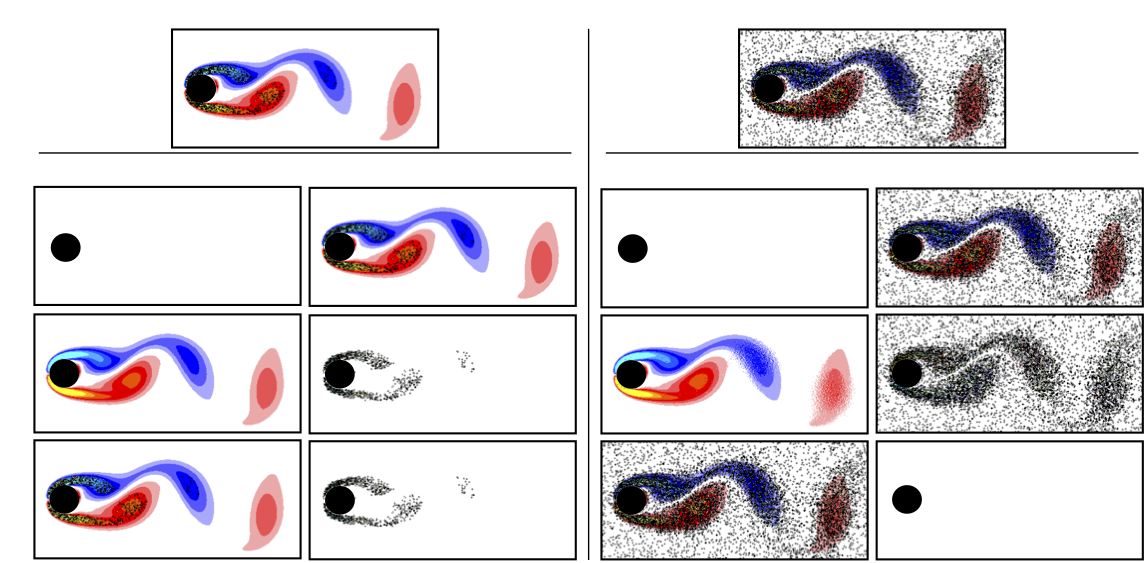}
\put(25, 47.25){${\bf X}$}
\put(74.5, 47.25){${\bf X}$}
\put(13.5, 33.3){${\bf L}$}
\put(38, 33.3){${\bf S}$}
\put(62.5, 33.3){${\bf L}$}
\put(88, 33.3){${\bf S}$}
\put(0,14){\begin{sideways}\textbf{$\lambda$} = 1\end{sideways}}
\put(0,24){\begin{sideways}\textbf{$\lambda$} = 0.1\end{sideways}}
\put(0,2.25){\begin{sideways}\textbf{$\lambda$} = 10\end{sideways}}
\end{overpic}
\vspace{-.075in}
\caption{RPCA filtering removes vorticity-biased corruption from simulated flow past a cylinder at Reynolds number $100$.  Unlike results shown in Fig.~\ref{Fig:CylResults}, corrupt entries are concentrated in regions of high vorticity instead of being uniformly distributed.  In the flow on the left, $\eta=1\%$ of the velocity field measurements are corrupted and on the right $\eta=10\%$ of the velocity field measurements are corrupted. All frames show resultant vorticity fields.}\label{Fig:VortSkew}
\end{center}
\end{figure}

\begin{figure}
\vspace{-.075in}
\begin{center}
\begin{overpic}[width=.98\textwidth]{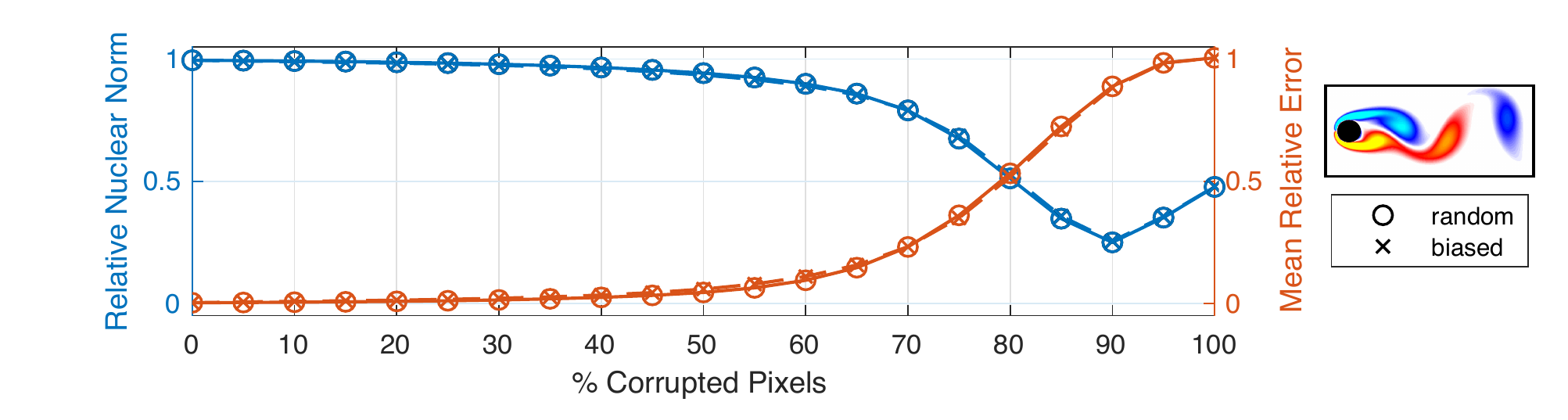}
\end{overpic}
\vspace{-.05in}
\caption{Error 
($\boldsymbol{||X_{uncorrupted} - L||_F/||X_{uncorrupted}||_F}$) 
and relative nuclear norm ($\boldsymbol{||L||_*/||X||_* 
= sum(\sigma_L)/sum(\sigma_{X,uncorrupted})}$)
of the low rank matrix ${\textbf{L}}$ 
compared with the uncorrupted data 
${\textbf{X}}$ for varying percentages of corruption.}
\label{Fig:Statistics}
\end{center}
\end{figure}

Figure~\ref{Fig:VortSkew} shows the results of RPCA filtering on the simulated data for the second case of vorticity-biased corruption.  
Again, in all cases, the theoretically optimal value of $\lambda=1$ yields the best segmentation of the corruption into the matrix ${\bf S}$. 
When the rate of corruption is increased from $1\%$ (left) to $10\%$ (right), the free-stream flow begins to become corrupted, resembling the uniform corruption case in Fig.~\ref{Fig:CylResults}.  
The mean error and relative nuclear norm of the low-rank matrix (${\bf L}$) compared to the true, uncorrupted data (${\bf X}$) are shown in Fig.~\ref{Fig:Statistics} for varying percentages of corrupt entries. Statistically, results are similar for vorticity-biased and randomly-distributed corruption.  
In both cases, RPCA filtering is remarkably robust to corruption, even for corruption in excess of $50\%$ of the measurements.  
This laminar vortex shedding example is an ideal application for RPCA filtering, as the true flow field is low rank and the corruption is sparse; it is unlikely that this will hold as well for data exhibiting broadband turbulence.
%
%
%

\subsubsection{PCA analysis for cylinder wake flows}
\begin{figure}
\vspace{.075in}
\begin{center}
\begin{overpic}[width=.934\textwidth]{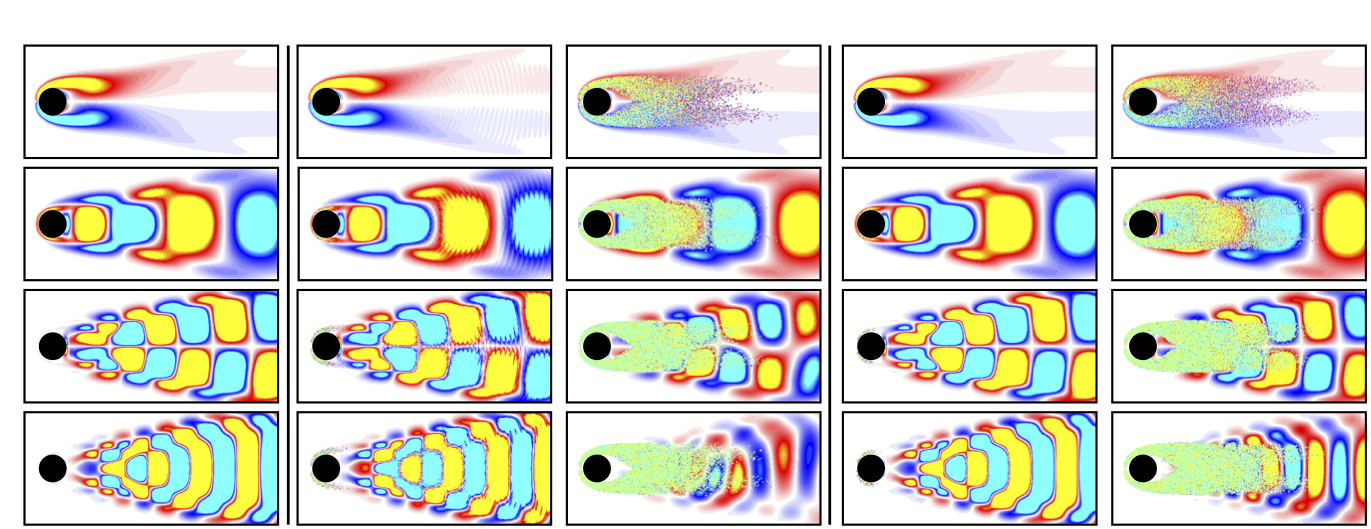}

\put(5,36.5){\textbf{True modes}}
\put(22,36.5){\textbf{RPCA (2 cycles)}}
\put(43,36.5){\textbf{PCA (2 cycles)}}

\put(62,36.5){\textbf{RPCA (5 cycles)}}
\put(83,36.5){\textbf{PCA (5 cycles)}}

\put(-3.5,15){\begin{sideways}\textbf{mode}\end{sideways}}
\put(0,30.5){\textbf{1}}
\put(0,21.5){\textbf{3}}
\put(0,12.5){\textbf{5}}
\put(0,3.5){\textbf{7}}

\end{overpic}
\vspace{-.075in}
\caption{Odd PCA vorticity  modes of the cylinder simulations from Fig.~\ref{Fig:VortSkew} with 1\% of velocity measurements corrupted with a bias towards regions of high vorticity.}\label{Fig:VortSkewModes1}
\vspace{.15in}
\end{center}
\vspace{-.15in}
\end{figure}

\begin{figure}
\vspace{.075in}
\begin{center}
\begin{overpic}[width=.934\textwidth]{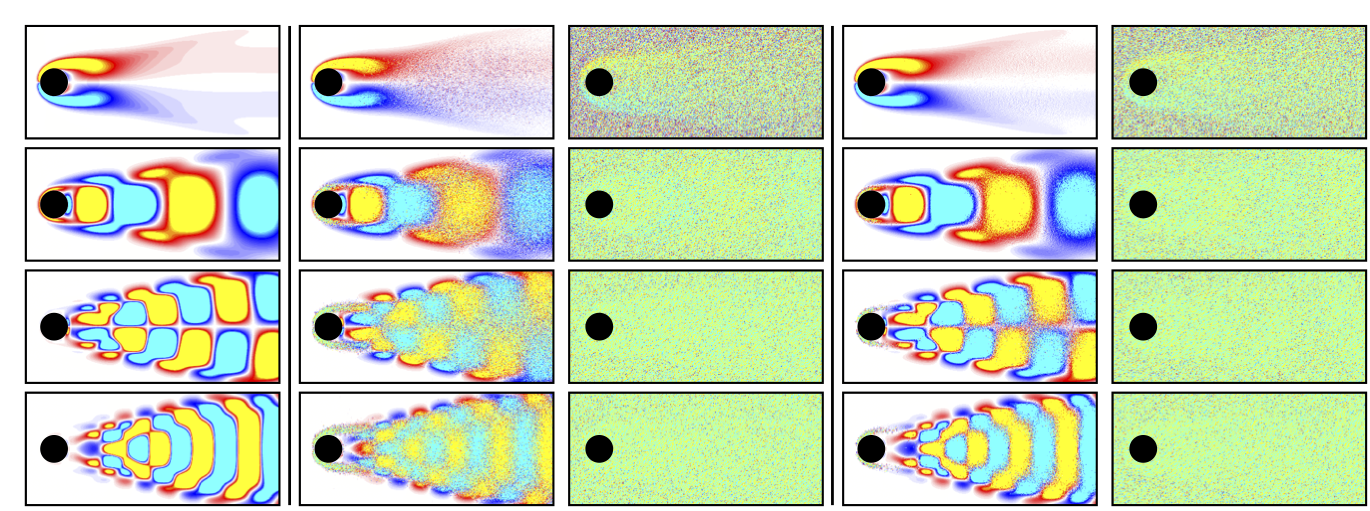}

\put(5,36.5){\textbf{True modes}}
\put(22,36.5){\textbf{RPCA (2 cycles)}}
\put(43,36.5){\textbf{PCA (2 cycles)}}

\put(62,36.5){\textbf{RPCA (5 cycles)}}
\put(83,36.5){\textbf{PCA (5 cycles)}}

\put(-3.5,15){\begin{sideways}\textbf{mode}\end{sideways}}
\put(0,30.5){\textbf{1}}
\put(0,21.5){\textbf{3}}
\put(0,12.5){\textbf{5}}
\put(0,3.5){\textbf{7}}

\end{overpic}
\vspace{-.075in}
\caption{Odd PCA vorticity modes of the cylinder simulations from Fig.~\ref{Fig:VortSkew} with 10\% of velocity measurements corrupted with a bias towards regions of high vorticity.}\label{Fig:VortSkewModes10}
\end{center}
\vspace{-.15in}
\end{figure}

We now investigate the impact of RPCA filtering on modal decompositions.  
Figures~\ref{Fig:VortSkewModes1} and~\ref{Fig:VortSkewModes10} show four leading PCA and RPCA modes for $1\%$ and $10\%$ vorticity-biased corruption, respectively.  
The first mode corresponds to the mean flow, and the remaining modes come in energetic pairs where the corresponding coefficients $\alpha_{2j}$ and $\alpha_{2j+1}$ oscillate sinusoidally at the same frequency but $\pi/2$ out of phase, sweeping out a circle in the phase plane.  
Thus, we only show one mode, $\boldsymbol{u}_{2j+1}$ from each of the first three energetic mode pairs.  
In all cases, the RPCA modes show dramatic improvement, while significant artifacts remain in the PCA modes.  
We also investigate the effect of increasing the amount of data, and there is a clear improvement in RPCA modes from 2 to 5 vortex shedding cycles; in contrast, the PCA modes do not improve appreciably with more data.  

\begin{figure}
\vspace{.075in}
\begin{center}
\begin{overpic}[width=.934\textwidth]{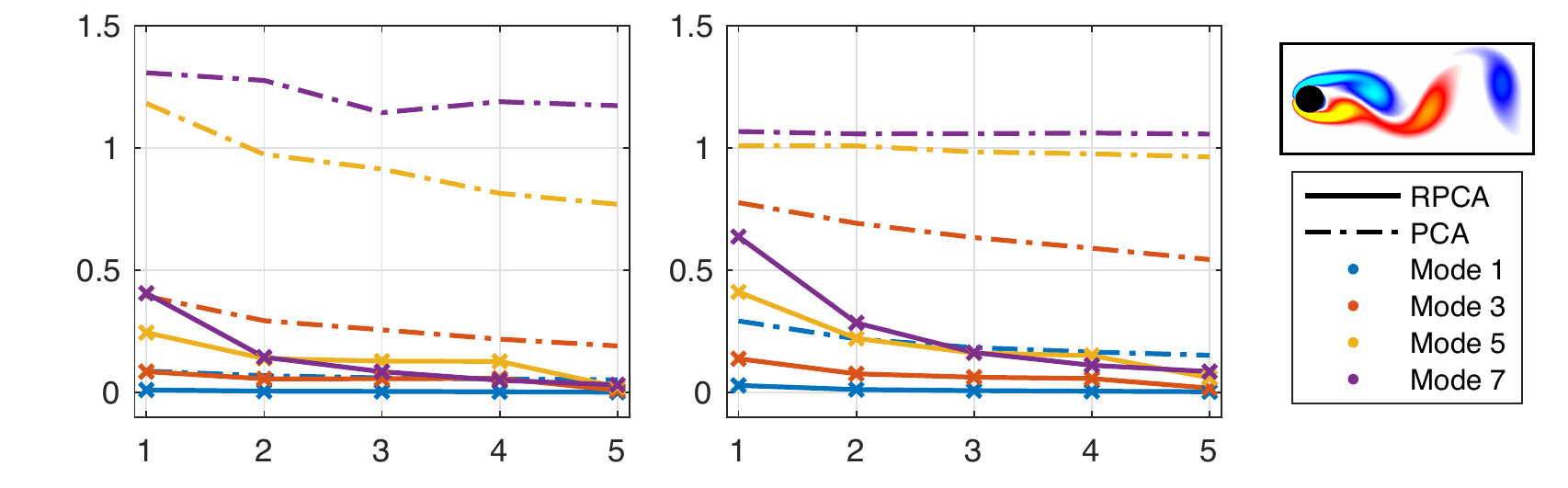}
\put(21.5,31){\textbf{$\eta = 1\%$}}
\put(58.5,31){\textbf{$\eta = 10\%$}}
\put(20.25,-0.75){\textbf{periods}}
\put(58,-0.75){\textbf{periods}}
\put(-0,6.5){\begin{sideways}\textbf{$||U_{\text{i, true}} - U_{\text{i,RPCA}}||_{2}$}\end{sideways}}

\end{overpic}
\caption{$L_2$ error between the true PCA modes of the clean cylinder simulation data ${\bf X}$ and the RPCA and PCA modes for corrupted data with $1\%$ (left) and $10\%$ (right) vorticity-biased corruption. }\label{Fig:L2CylSim}
\end{center}
\end{figure}

To quantify the improvement observed above, we compute the $L_2$ error between the PCA and RPCA modes of corrupted data and the PCA modes for the clean data (i.e., DNS results) as a function of the number of shedding periods. As show in Fig.~\ref{Fig:L2CylSim}, the RPCA mode velocity fields quickly converge to a small error as the amount of data is increased for both the $1\%$ and $10\%$ corruption cases, while the PCA modes converge much more slowly and still have considerable error after 5 shedding periods are included in the analysis.

\begin{figure}
\vspace{.075in}
\begin{center}
\begin{overpic}[width=.934\textwidth]{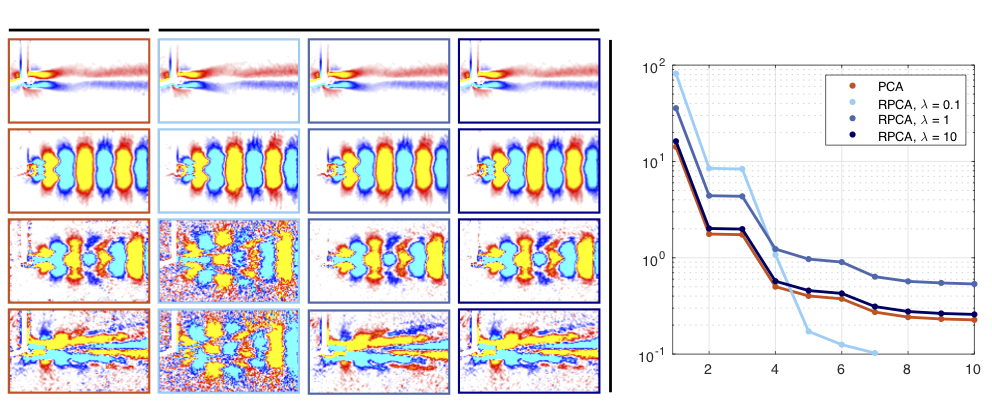}
\put(5, 40){\textbf{PCA}}
\put(35,40){\textbf{RPCA}}

\put(19,-0.5){\textbf{$\lambda = 0.1$}}
\put(35.5,-0.5){\textbf{$\lambda = 1$}}
\put(50,-0.5){\textbf{$\lambda = 10$}}

\put(-3.5,16.5){\begin{sideways}\textbf{mode}\end{sideways}}
\put(-1,32.5){\textbf{1}}
\put(-1,23.5){\textbf{3}}
\put(-1,14.5){\textbf{5}}
\put(-1,6){\textbf{7}}

\put(62.5,14.5){\begin{sideways}\textbf{\% Energy}\end{sideways}}
\put(82,1){\textbf{mode}}

\end{overpic}
\vspace{.075in}
\caption{Odd PCA vorticity modes of the experimental cylinder data for PCA and RPCA at $\lambda = 0.1, 1,$ and $10,$ along with their singular values.}\label{Fig:PIVmodes}
\end{center}
\vspace{-.15in}
\end{figure} 

The modes for the PIV data for the cylinder flow are shown in Fig.~\ref{Fig:PIVmodes}.  
This figure highlights the effect of $\lambda$, the sparsity hyperparameter, which was previously discussed with respect to Fig.~\ref{Fig:CylResults}.  
In this case, we do not have a clean ground-truth data set to compare against.  
Although the flow field in Fig.~\ref{Fig:CylResults} appears to have less corruption for $\lambda=0.1$, here we see that all RPCA modes after the first three modes are heavily filtered, as seen in the rapid drop off in the singular values after the third mode.  The corresponding modes are highly corrupt, further supporting that $\lambda=0.1$ is not a good choice.  
In contrast, the RPCA modes for the theoretically optimal $\lambda=1$ case appear to have slightly less free-stream corruption than the PCA modes.  
Also, as expected, for a large enough value of $\lambda$, the RPCA filtering has little effect on the modes. 

%

\begin{figure}
\vspace{.075in}
\begin{center}
\begin{overpic}[width=.934\textwidth]{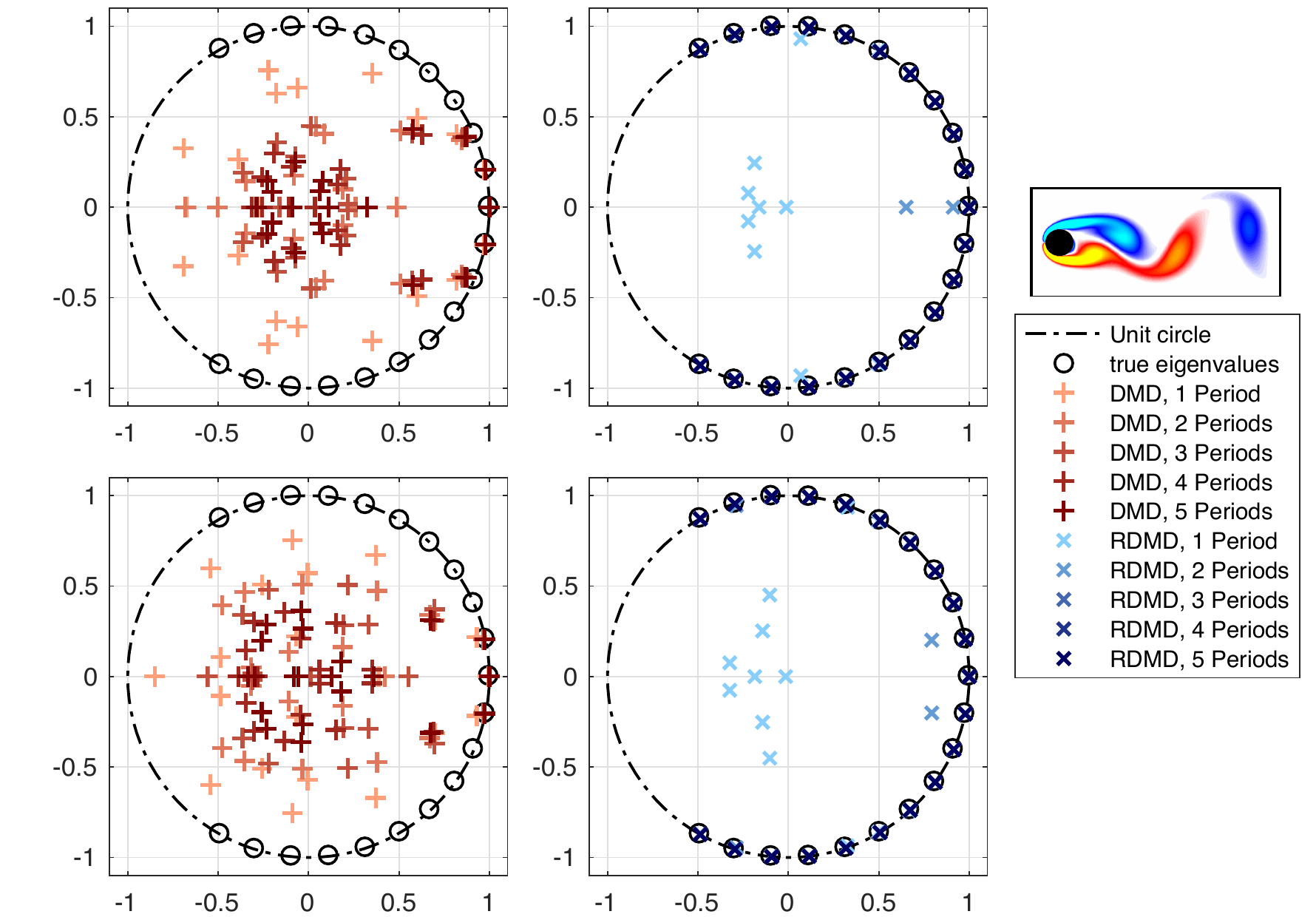}

\put(21,70.5){\textbf{DMD}}
\put(56,70.5){\textbf{RDMD}}
\put(-1.5,50.5){\begin{sideways}\textbf{$\eta = 1\%$}\end{sideways}}
\put(-1.5,15){\begin{sideways}\textbf{$\eta = 10\%$}\end{sideways}}
\put(21,-2){\textbf{$Re(\gamma)$}}
\put(57.5,-2){\textbf{$Re(\gamma)$}}

\put(2,15.5){\begin{sideways}\textbf{$Im(\gamma)$}\end{sideways}}
\put(2,51.5){\begin{sideways}\textbf{$Im(\gamma)$}\end{sideways}}

\end{overpic}
\vspace{.075in}
\caption{Discrete-time DMD eigenvalues for the simulated cylinder data for small and large amounts of corruption and for increasing amounts of training data.  In all cases, the RPCA-filtered DMD results dramatically outperform the standard DMD results.} \label{Fig:eigvals}
\end{center}
\vspace{-.15in}
\end{figure}

\subsubsection{DMD analysis for cylinder wake flows}
\begin{figure}
\begin{center}
\begin{overpic}[width=.934\textwidth]{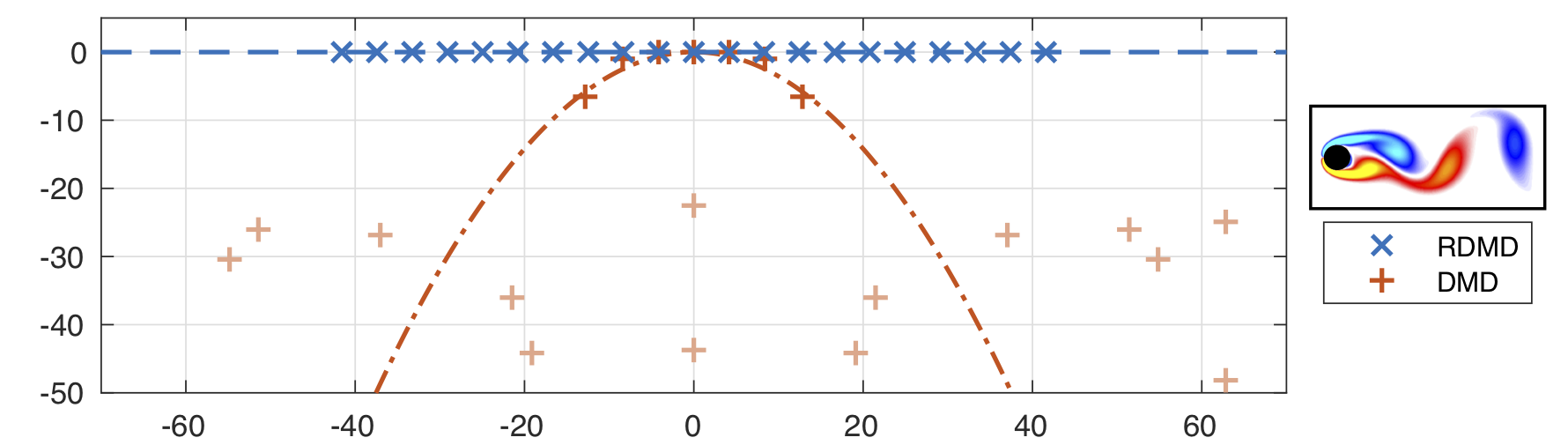}

\put(42,-2){\textbf{$Im(\omega)$}}

\put(-1,12){\begin{sideways}\textbf{$Re(\omega)$}\end{sideways}}
\end{overpic}
\vspace{.075in}

\caption{Continuous-time DMD eigenvalues for the simulated cylinder data along with parabolic eigenvalue fits to estimate the error as in~\cite{Bagheri2014pof}. Here we use $5$ vortex shedding periods with $\eta=1\%$ corrupt values. The RDMD parabolic coefficient is approximately $2 \times 10^4$ times smaller than the DMD coefficient.} \label{Fig:eigvals_cont}
\end{center}
\vspace{-.15in}
\end{figure}

\begin{figure}
\vspace{-.075in}
\begin{center}
\begin{overpic}[width=.934\textwidth]{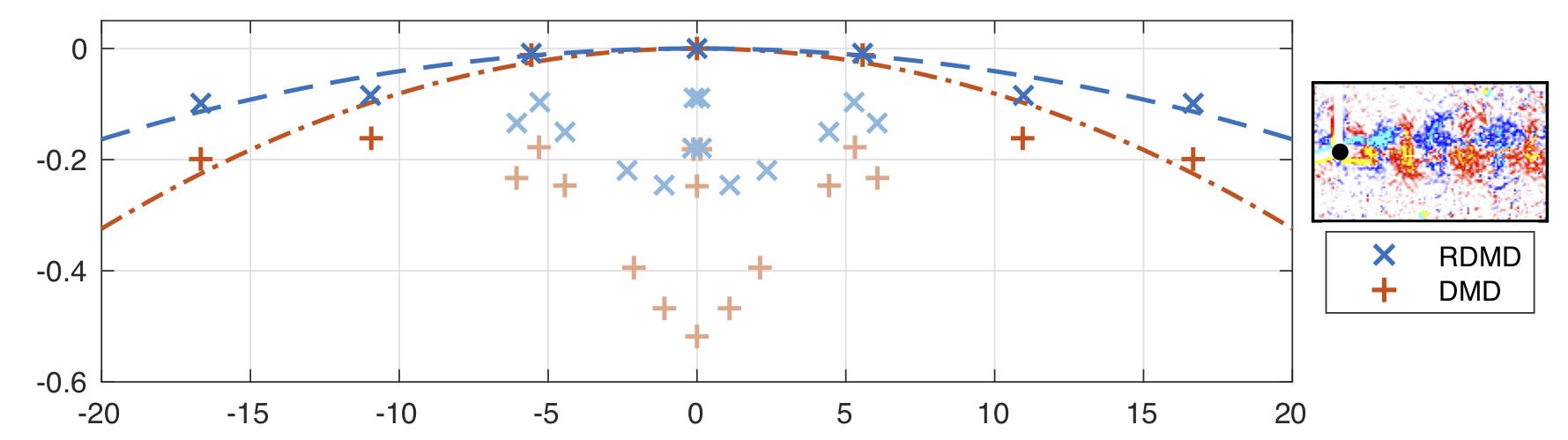}
\put(42,-2){\textbf{$Im(\omega)$}}
\put(-1,12){\begin{sideways}\textbf{$Re(\omega)$}\end{sideways}}
\end{overpic}
\vspace{.075in}
\caption{Continuous-time DMD eigenvalues for the PIV cylinder data along with parabolic eigenvalue fits to estimate the error as in~\cite{Bagheri2014pof}. The RDMD parabolic coefficient is approximately two times smaller than the DMD coefficient.}\label{Fig:eigvals_piv_cont}
\end{center}
\vspace{-.15in}
\end{figure}

DMD is known to be quite sensitive to noisy data, making this a challenging test case for RPCA filtering.  
Figure~\ref{Fig:eigvals} shows the discrete-time eigenvalues for the cylinder DNS data with vorticity-biased corruption.  
For the cylinder wake, the uncorrupted or true DMD eigenvalues may be computed from the noiseless data, and they are equally spaced on the unit circle in the complex plane.  
In all cases, the RPCA-filtered DMD (RDMD) data results in dramatically better agreement with the uncorrupted or true DMD eigenvalues compared with the corrupted DMD eigenvalues.  Even with only a single period of data and $\eta=10\%$ corruption, the RDMD values capture the first six low-frequency mode pairs; in contrast, even with five periods of data and as little as $\eta=1\%$ corruption, corrupted DMD only captures the first two low-frequency mode pairs, and with considerably more spurious damping.  
To see this more clearly, we plot the eigenvalues in continuous-time in Fig.~\ref{Fig:eigvals_cont}, where the $x$-axis is the imaginary eigenvalue component and the $y$-axis is the real eigenvalue component, which is a standard way to plot DMD eigenvalues~\cite{Schmid2010jfm}.  
Here, the best-fit parabolas for the RDMD eigenvalues and the first seven corrupted DMD eigenvalues are shown in dashed lines.   
The curvature of these parabolas is directly related to the noise amplitude, as in Eq.~\eqref{Eq:Shervin} from~\cite{Bagheri2014pof}.  
The same continuous-time eigenvalue plot is shown for the PIV cylinder wake data in Fig. \ref{Fig:eigvals_piv_cont}.  
In both cases, we see that the parabolic eigenvalue fit for the RDMD eigenvalues has a smaller curvature than for the corrupted DMD eigenvalues, indicating a quantitative and significant reduction in noise.

%

\subsection{Turbulent channel flow} 
\begin{figure}
\vspace{.1in}
\begin{center}
\begin{overpic}[width=.98\textwidth]{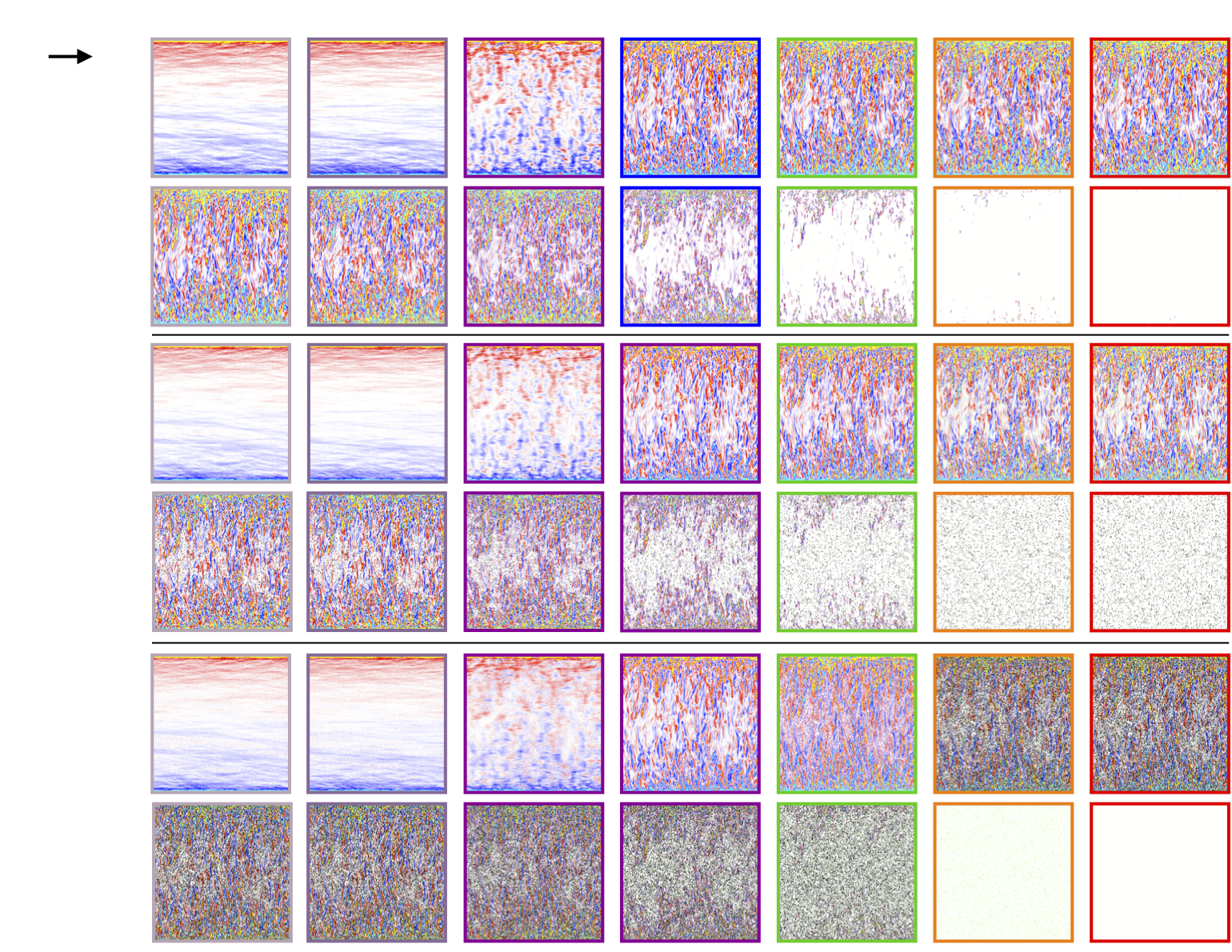}
\put(9,68){\textbf{L}}
\put(9,43){\textbf{L}}
\put(9,18.5){\textbf{L}}
\put(9,57){\textbf{S}}
\put(9,32){\textbf{S}}
\put(9,6.5){\textbf{S}}

\put(3,58.5){\begin{sideways}$\eta = 0\%$\end{sideways}}
\put(3,33.5){\begin{sideways}$\eta = 2\%$\end{sideways}}
\put(3,9){\begin{sideways}$\eta = 10\%$\end{sideways}}

\put(14,75){$\lambda = 0.1$}
\put(26.25,75){$\lambda = 0.2$}
\put(40,75){$\lambda = 0.5$}
\put(53,75){$\lambda = 1$}
\put(65.5,75){$\lambda = 2$}
\put(78,75){$\lambda = 5$}
\put(90,75){$\lambda = 10$}
\put(2,74){$U_{\infty}$}

\end{overpic}
\vspace{-.1in}
\caption{RPCA filtering for turbulent channel flow vorticity fields with various levels of added noise and tuning parameter $\lambda$.   $\eta$ represents the percentage of corrupted measurements in the velocity fields. The border colors match the color of the curve at the corresponding $\eta$ in Fig.~\ref{Fig:TurbSpec}.}\label{Fig:Channel}
\end{center}
\end{figure}

\begin{figure}
\begin{center}
\begin{overpic}[width=1\textwidth]{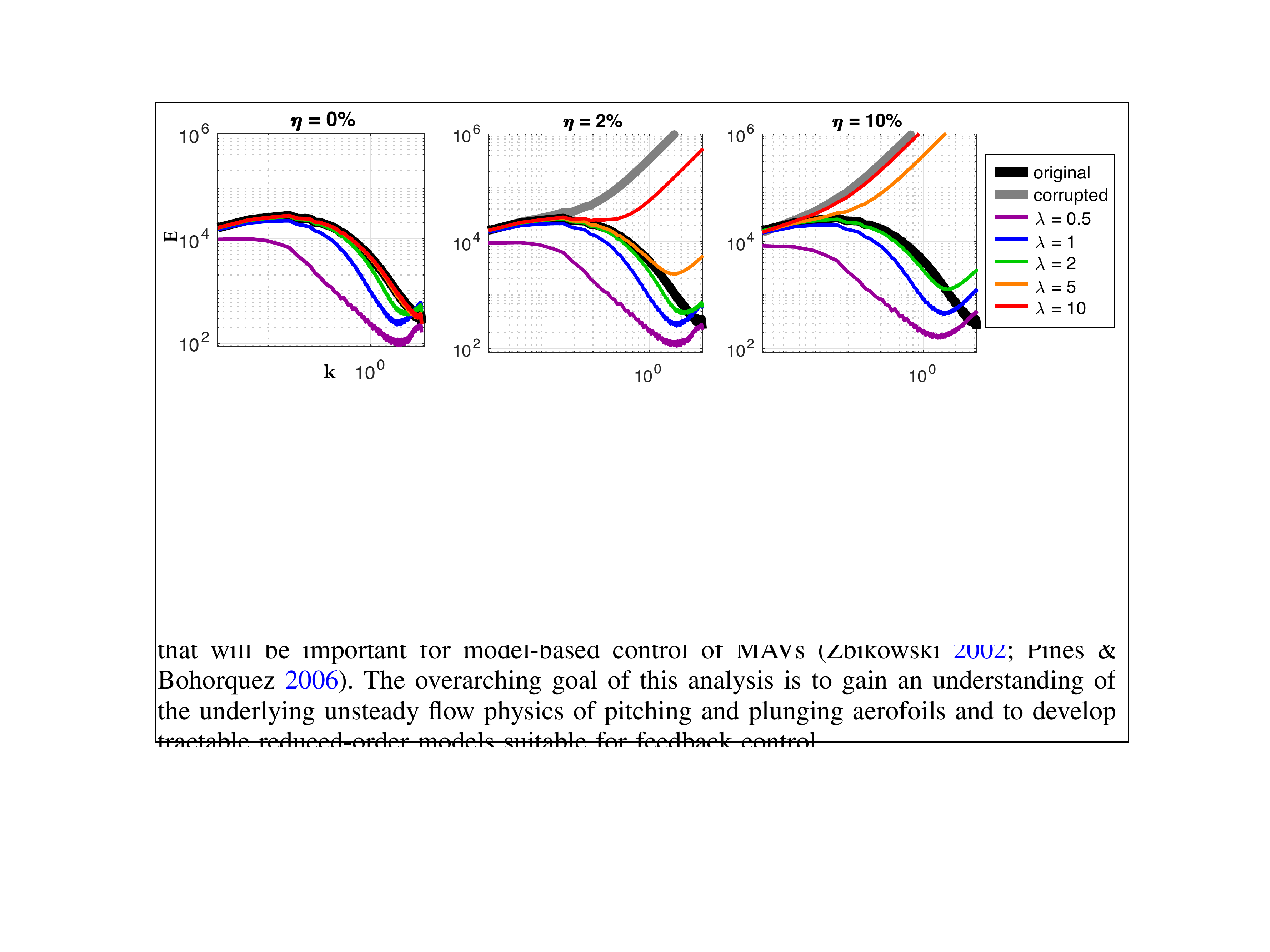}

\end{overpic}
\caption{Turbulent kinetic energy (TKE) spectra for various levels of corruption and RPCA filtering. The TKE profiles provide a summary of the filtering that occurs at various scales. As corruption increases, the filtering remove more high-frequency information.}\label{Fig:TurbSpec}
\end{center}
\end{figure}

The tradeoff between filtering corruption and small-scale structures is also apparent in the turbulent channel flow DNS.  
Unlike the cylinder wake, this flow field contains broadband turbulent phenomena across multiple spatial and temporal scales.  
Figure~\ref{Fig:Channel} shows RPCA filtering for various levels of corruption, sweeping across the tuning parameter $\lambda$.  
The corresponding turbulent kinetic energy (TKE) is shown in Fig.~\ref{Fig:TurbSpec}, providing a summary of the various scales that are filtered.  
The value of $\lambda$ that preserves the true TKE spectrum varies with the degree of velocity field corruption. 
In the uncorrupted case ($\eta=0$), we can clearly see the effect of filtering on the turbulent coherent structures, indicating that some fine-scale structures are filtered for $\lambda=1$.  
As the degree of corruption increases to $\eta=2\%$, we see that the curves for $\lambda \leq 2$ remain relatively unchanged, although the $\lambda > 2$ curves begin to exhibit spurious high-frequency spatial structures (i.e., corruption is present in $\bf L$).  
As the rate of corruption increases to $\eta=10\%$, spurious high-frequency energy also appears for $\lambda=2$.  
In this case, it is clear that the optimal filtering value $\lambda$ changes with the level of corruption.  
For relatively limited corruption, a larger value of $\lambda$ may be used, but must be decreased towards the theoretically optimal value of $\lambda=1$ for higher levels of corruption. 
Finally, we note that, unlike the cylinder wake cases, it is not surprising that $\lambda = 1$ is sub-optimal because the channel flow is not fundamentally low rank which deviates from an underlying assumption of the RPCA algorithm.

\subsection{Cross-flow turbine wake} 
\begin{figure}
\vspace{.05in}
\begin{center}
\begin{overpic}[width=.965\textwidth]{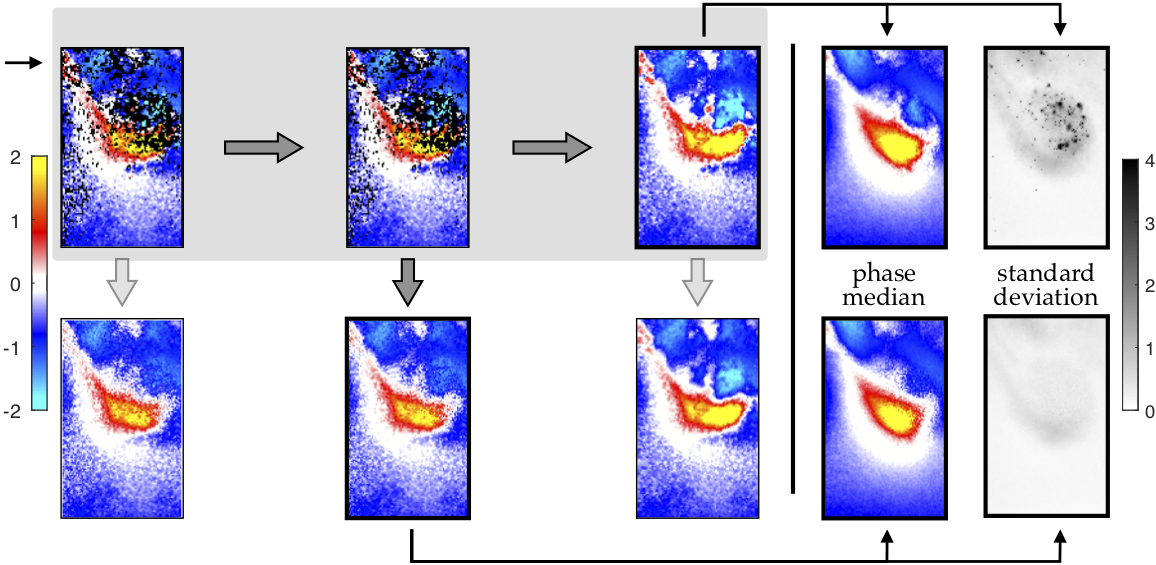}



%
%



\put(0,45){$U_{\infty}$}
\put(5.5,45.7){\textbf{Standard PIV Processing Pipeline}}
\put(37.5,23.5){\textbf{RPCA}}


\put(41.5,38.5){\small interpolation}

\put(19.5,40.5){\small vector}
\put(17.5,38.5){\small validation}


\put(-2,15){\begin{sideways}U-Velocity (m/s)\end{sideways}}
\put(101.5,11){\begin{sideways}Standard Deviation (m/s)\end{sideways}}

\end{overpic}
\caption{RPCA filtering of cross-flow turbine wake PIV data.  The standard PIV processing pipeline (top row) includes several steps where RPCA filtering can be applied (bottom row).  In the cross-correlated streamwise velocity field (top left), $23\%$ of the velocity vectors are missing.  Vector validation reduces the missing vectors to $20\%$.  Finally, linear interpolation is used to fill in these missing vectors.  
In all cases, RPCA filtering captures the relevant phase-averaged coherent structures with fewer outliers and missing data, which appear as dark spots in the standard deviation plot.  
}\label{Fig:CrossFlowResults}
\end{center}
\end{figure}

As a final example, we consider the use of RPCA filtering to identify outliers and fill in missing PIV data collected in the wake of a cross-flow turbine, as shown in Fig.~\ref{Fig:CrossFlowResults}.   
There are several stages in the PIV processing pipeline where RPCA filtering could be applied, including after initial cross-correlation, after conventional normalized median filter vector validation, and after linear interpolation. For the cases shown here, we use $\lambda =1.6$, which results in a velocity in the bypass flow, or lower third of the frame, that visually matches the frequency content of the unfiltered data.   
There are enough missing velocity vectors (23\% and 20\%, respectively) to degrade the effectiveness of both median filtering and interpolation.  
In contrast, RPCA filtering produces flow fields that capture dominant coherent structures for either cross-correlated or median-filtered fields. 
Finally, by investigating the standard deviation of all flow fields collected at a given turbine angular position (i.e., phase), it is clear that the RPCA filtering can be used to remove artifacts introduced by linear interpolation. 
This is consistent with the intuition that vector validation and interpolation should fail in these regions where there is high density of missing data is spatially clustered.  

\begin{figure}
\begin{center}
\begin{overpic}[width=.925\textwidth]{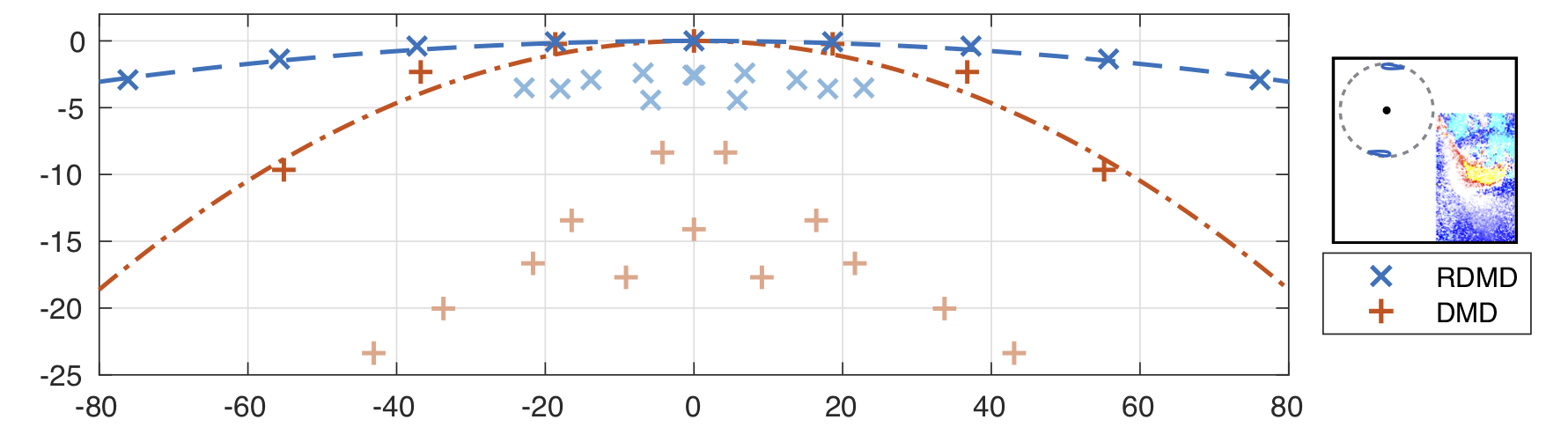}
\put(42,-.5){\textbf{$Im(\omega)$}}
\put(-1,12){\begin{sideways}\textbf{$Re(\omega)$}\end{sideways}}
\end{overpic}
\vspace{-.075in}
\caption{Continuous-time DMD eigenvalues for the turbine wake PIV data, along with parabolic eigenvalue fits to estimate the error as in~\cite{Bagheri2014pof}. The RDMD parabolic coefficient is approximately six times smaller than the DMD coefficient.  
} \label{Fig:eigvals_turbine_cont}
\end{center}
\vspace{-.15in}
\end{figure} 

The continuous-time DMD eigenvalues for the cross-flow turbine wake are shown in Fig.~\ref{Fig:eigvals_turbine_cont}. In this plot, the parabolic fits for DMD eigenvalues computed after interpolation and RDMD eigenvalues computed after vector validation are displayed as dotted lines. For this case, the coefficient of the parabolic fit for the RDMD-based eigenvalues is six times smaller than the parabolic fit coefficient for the DMD-based eigenvalues. 
This demonstrates a significant quantitative improvement of the DMD spectrum using RPCA filtering to process the data.

\section{Conclusions and discussion}\label{Sec:Conclusions}
In this work, we have demonstrated the ability of RPCA filtering to effectively recover dominant coherent structures from corrupt flow fields with missing measurements.  
Unlike standard POD/PCA, which is based on least squares and is susceptible to outliers and corruption, RPCA utilizes sparse optimization to separate a data matrix into a low-rank matrix containing correlated structures and a sparse matrix containing the spurious entries.

We apply RPCA filtering to several types of fluid flow data (DNS and PIV), ranging from laminar vortex shedding behind a circular cylinder, to fully turbulent channel flow DNS, and concluding with an experimental flow past a cross-flow turbine.  
These flows exhibit a variety of phenomena and a range of measurement quality.  
The DNS examples provide us with a baseline, where it is possible to add corruption to quantitatively assess the performance of RPCA.  
For flow past a cylinder in DNS, RPCA filtering is extremely effective at separating the true flow field from considerable corruption, with robust recovery even in flow fields with excess of 50\% of the measurements corrupted.  
In the experimental counterpart, RPCA is still able to remove large outliers and corruption, although there is a tradeoff between filtering the background turbulence and coherent structures in the wake.  
The fully turbulent channel flow DNS provides an opportunity to more fully explore this tradeoff in a controlled setting, where we can incrementally increase the corruption ratio and observe the filtering effects on various spatial frequencies.  
As expected, an increasingly aggressive filtering leads to degradation at higher wavenumbers, although dominant coherent structures are robustly preserved.  
Finally, the wake behind a cross-flow turbine provides a practical real-world flow that directly benefits from improved PIV processing.  
In all three wake flows we also assess the performance of RPCA filtering to yield more accurate modal decompositions.  
Although we do not have ground truth measurements and modal decompositions, except in the case of direct numerical simulations, we know that continuous-time DMD eigenvalues should be arranged on the imaginary axis in the complex plane for clean data, and deviations from this may be quantified using the derivation from Bagheri~\cite{Bagheri2014pof}. In all three cases, we see considerable reduction in spurious damping, indicating the de-noising effectivness of RPCA. 
Based on these results, we believe that RPCA can be a valuable algorithm in the arsenal of PIV processing and filtering techniques, particularly when the processing pipeline culminates in modal analysis.  

There are a number of future directions motivated by this work.  
First, RPCA depends on the hyperparameter $\lambda$, and a better understanding of how to objectively choose $\lambda$ for different conditions is important.  
Because RPCA is based on sparse, non-convex optimization, it is also likely that improved optimization techniques may improve speed and robustness.  
Although this work considered three-dimensional flows, the data comprised two-dimensional cross-sections, and the current analysis could be extended to flow volumes.  
In principle, the RPCA method should generalize, although there may be computational scaling challenges.  
It would also be useful to extend this work to PIV measurements of other turbine configurations~\citep{posa2016wake}. 
Finally, the quality of the RPCA filtered flow fields for additional downstream analyses should be assessed for example, 
in dynamical systems modeling via Galerkin projection~\citep{Noack2003jfm} or regression~\citep{Loiseau2017jfm} onto the filtered modes and in control~\citep{berger2015active}. 

\section*{Acknowledgments}
\noindent We gratefully acknowledge funding from the Army Research Office (ARO W911NF-19-1-0045) and Naval Facilities Engineering Command.  We would also like to thank Jared Callaham, Nathan Kutz, Kazuki Maeda, and Joshua Proctor for valuable discussions.

\setlength{\bibsep}{3.5pt plus 1ex}
\begin{spacing}{.01}
\small
\bibliographystyle{plain}

\end{spacing}

\end{document}